\documentclass[aps,prd,nofootinbib,eqsecnum,showkeys,tightenlines,10pt,superscriptaddress,preprintnumbers]{revtex4-2}

\usepackage{amsmath,amssymb,mathtools,mathrsfs,stix,slashed,cancel}

\usepackage{verbatim}     
\usepackage{ulem}         
\usepackage{comment}
\usepackage{xcolor}

\usepackage{graphicx}
\graphicspath{{images/}}
\usepackage{subfig}
\usepackage{caption}
\usepackage{float}
\usepackage{multirow}

\usepackage[
colorlinks=true,        
citecolor=blue,        
linkcolor=blue,         
urlcolor=blue           
]{hyperref}  

\usepackage{orcidlink}

\begin{document}

\preprint{NU-QG-9}
\preprint{RUP-25-19}
\preprint{KUNS-3070}

\title{Primordial Black Hole Formation and Spin in Matter Domination Revisited}

\author{Weitao Ye \orcidlink{0009-0006-5981-1037}}
\email{weitao.ye@hotmail.com}
\affiliation{School of physics, Huazhong University of Science and Technology, Wuhan 430074, P. R. China}

\author{Yungui Gong \orcidlink{0000-0001-5065-2259}}
\email{gongyungui@nbu.edu.cn}
\affiliation{Institute of Fundamental Physics and Quantum Technology, Department of Physics, School of Physical Science and Technology, Ningbo University, Ningbo, Zhejiang 315211, China}

\author{Tomohiro Harada \orcidlink{0000-0002-9085-9905}}
\email{harada@rikkyo.ac.jp}
\affiliation{Department of Physics, Rikkyo University, Toshima, Tokyo 171-8501, Japan}

\author{Zhaofeng Kang}
\email{zhaofengkang@gmail.com}
\affiliation{School of physics, Huazhong University of Science and Technology, Wuhan 430074, P. R. China}

\author{Kazunori Kohri \orcidlink{0000-0003-3764-8612}}
\email{kazunori.kohri@gmail.com}
\affiliation{Division of Science, National Astronomical Observatory of Japan, 2-21-1 Osawa, Mitaka, Tokyo 181-8588, Japan}
\affiliation{Theory Center, IPNS, High Energy Accelerator Research Organization (KEK), 1-1 Oho, Tsukuba, Ibaraki 305-0801, Japan}
\affiliation{School of Physical Sciences, Graduate University for Advanced Studies (SOKENDAI), 2-21-1 Osawa, Mitaka, Tokyo 181-8588, Japan}
\affiliation{Kavli IPMU (WPI), UTIAS, The University of Tokyo, Kashiwa, Chiba 277-8583, Japan}

\author{Daiki Saito  \orcidlink{0000-0003-1624-9268}}
\email{saito@tap.scphys.kyoto-u.ac.jp}
\affiliation{Department of Physics, Kyoto University, Kyoto 606-8502, Japan}

\author{Chul-Moon Yoo \orcidlink{0000-0002-9928-4757}}
\email{yoo.chulmoon.k6@f.mail.nagoya-u.ac.jp}
\affiliation{
Graduate School of Science, Nagoya University, 
Nagoya 464-8602, Japan
}
\affiliation{Kobayashi-Maskawa Institute for the Origin of Particles and the Universe (KMI), Nagoya 464-8602, Japan}

\begin{abstract}
    In this article, we calculate the mass distribution of primordial black holes (PBHs) formed in the matter-dominated (MD) era by the peak theory. 
    We apply the Zel'dovich approximation to track the nonlinear evolution of overdensities and compute the PBH abundance and mass function by incorporating a PBH formation criterion based on the hoop conjecture.
    We find that the PBH abundance $\beta$ follows the scaling law $\beta \simeq A_\gamma \sigma_h^{*5}$ for $\sigma_h^*\ll 1$.
    Here, $\sigma_h^*$ is the quantity that characterizes the variance of the density fluctuation at the horizon entry.
    We also find that, in contrast to the previous estimates, the PBH spin is very small for $\sigma_h^*\ll 1$ but could be larger for larger $\sigma_h^*$ and broader power spectra.
    Finally, specializing to a monochromatic power spectrum, we prove analytically that the PBH mass distribution becomes effectively monochromatic and reveal that the resultant PBH abundance is approximately 19 times the previous prediction. 
\end{abstract}

\maketitle

\tableofcontents

\section{Introduction}

Primordial black holes (PBHs) were first proposed over 50 years ago \cite{Zeldovich:1967lct,Hawking:1971ei,Carr:1974nx}. Formed in the early universe, PBHs can possess a wide range of masses, distinguishing them from black holes resulting from stellar collapse. In recent years, PBHs have garnered significant attention, as they could potentially explain gravitational waves from binary black hole mergers \cite{Bird:2016dcv,Sasaki:2016jop,Clesse:2016vqa} and serve as viable dark matter candidates \cite{Carr:2020gox,Carr:2020xqk,Green:2020jor}. 
Although PBHs have not yet been directly detected, there are positive evidence for their existence, as summarized in the review \cite{Carr:2023tpt}.

In general, the formation mechanisms of PBHs are diverse, as summarized in a recent review \cite{Escriva:2022duf}.
The most extensively studied mechanism is the collapse of primordial overdense regions during the radiation-dominated (RD) era, leading to black hole formation. Typically, these overdense regions are considered to originate from quantum fluctuations during inflation. After inflation ends, these fluctuations re-enter the Hubble horizon and evolve accordingly. Due to the significant pressure in the RD era ($w=\frac{p}{\rho}=\frac{1}{3}$), only those density fluctuations that are sufficiently large at horizon entry, 
$\delta_h \coloneqq \frac{\rho - \rho_b}{\rho}|_h >\delta_{\rm th} \sim w$, can overcome pressure gradient forces and collapse into black holes.
Here, the subscript $h$ indicates that the fluctuation is evaluated at the time of horizon entry.
Since the amplitude of density fluctuations is generally assumed to follow a Gaussian probability distribution, the PBH formation probability is greatly suppressed to $\beta \simeq \sqrt{\frac{2}{\pi}} \exp \left(\frac{-\delta^2_{\rm th}}{2\sigma_h^2} \right) \ll 1$, where $\sigma_h\coloneqq \langle \delta^2_h \rangle^{1/2}$ is usually considered to be much less than 1. 
Moreover, according to the peak theory, for overdense regions with $\delta_h \gg \sigma_h$, the initial shape is nearly spherical
implying that PBHs formed during the RD era generally have low spin. 
For further details on PBHs in the RD era, see Refs.~\cite{Carr:1975qj,Niemeyer:1999ak,Shibata:1999zs,Musco:2004ak,Musco:2008hv,Musco:2012au,Harada:2013epa,DeLuca:2020bjf,Harada:2020pzb,Yoo:2024lhp}.

Prior to the RD epoch, an additional early matter dominated (EMD) era may have occurred, which can arise from the end of inflation~\cite{Albrecht:1982mp,Kofman:1994rk,Kofman:1997yn,Kawasaki:2000en,Amin:2014eta,Carr:2018nkm,Hasegawa:2020ctq,Aurrekoetxea:2023jwd,Padilla:2024iyr}, the domination of non-relativistic particles \cite{Khlopov:1980mg,polnarev1982era,Berlin:2016vnh,Tenkanen:2016jic,Allahverdi:2021grt}, or the presence of moduli fields in string theory constructions \cite{Green:1997pr,Kohri:2004qu,Kane:2015jia}.
In a pressureless environment ($w=0$), the collapse of density fluctuations into black holes is greatly enhanced, and consequently, the formation of PBHs during the matter-dominated (MD) era has attracted considerable interest in recent years. Although the threshold overdensity required to overcome pressure, $\delta_{\rm th}\sim w=0$, vanishes in this regime, PBH formation can still be suppressed by a variety of additional effects. 
In particular, anisotropic collapse \cite{Khlopov:1980mg,Harada:2016mhb}, 
spin \cite{Harada:2017fjm,Saito:2024hlj}, 
inhomogeneity \cite{Khlopov:1980mg,polnarev1982dustlike,Kokubu:2018fxy},
and velocity dispersion \cite{Harada:2022xjp} have been studied, each deepening our understanding of PBH formation during the MD era. However, these effects have typically been treated independently. In this work, we employ the peak theory to incorporate both anisotropic collapse and spin in a unified framework, yielding not only a more precise determination of the PBH formation probability but also the novel result that the spins of PBHs from the MD era are generally very small, which is contrary to previous expectations.

The importance of anisotropic collapse was first noted in Ref. \cite{Khlopov:1980mg} and subsequently refined by Ref.~\cite{Harada:2016mhb}.
Since gravitational collapse is the internal motion of an overdense region, tidal forces, which encoded in the tidal tensor (the second derivative of the gravitational potential with respect to coordinates), dominate the collapse. When the tidal forces are anisotropic, the overdense region collapses at different rates along different directions, leading not to a point‑like concentration but rather to a "pancake" configuration. Only if the anisotropy of the tidal field is sufficiently small can a black hole form. Ref. \cite{Harada:2016mhb} employed the Zel'dovich approximation \cite{Zeldovich:1969sb} to study the collapse process of the overdense region, and adopted the hoop conjecture \cite{Klauder:1972je,Misner:1973prb}
-- which states that a black hole forms if and only if the circumference $C$ of the pancake satisfies  $C\lesssim 4\pi M$ (with $M$ the mass of the region) -- as the formation criterion. By combining this with the Doroshkevich distribution (the probability distribution of the eigenvalues of the rescaled tidal tensor) \cite{1970Ap......6..320D}, they acquired the formation probability $\beta\simeq 0.056 \sigma_h^5$ for $\sigma_h \ll 1$. Because this scaling is a steep power law, it represents a much weaker suppression than the Gaussian cutoff characteristic of the RD era. However, they treated the initial shape of the overdense regions of the growth simply as spherical, neglecting deviations from sphericity that are generically present unless the density perturbation power spectrum is strictly monochromatic.
In reality, non‑sphericity affects PBH formation in two key ways: (1) by modifying the pancake circumference $C$, and (2) via tidal torques (the off‑diagonal components of the tidal tensor), which impart angular momentum to the collapsing region.

The dimensionless spin is defined by $\tilde{a}\coloneqq \frac{S}{M^2}$ and black holes must satisfy the Kerr bound $\tilde{a}_{\rm BH}<1$. The spin of PBHs formed during the MD era was first considered by Ref.~\cite{Harada:2017fjm}, and was recently revisited by Ref.~\cite{Saito:2024hlj} within the peak theory framework without assuming initial spherical symmetry. According to their analysis, for the vast majority of overdense regions, tidal torques can induce large angular momenta; concretely, for $\sigma_h \ll 1$, the average spin of collapsing regions behaves as  $\langle \tilde{a} \rangle \sim \sigma_h^{-1/2} \gg 1$, so that only a small subset which is denoted by $\beta_{\rm spin}$ satisfies $\tilde{a}<1$. Then, assuming that PBHs form with mean spin $\langle \tilde{a} \rangle_{\rm BH} \sim 1$,
they argued that the overall formation probability is further suppressed to $\beta = 0.056 \sigma_h^5 \times \beta_{\rm spin}$. However, they did not rigorously identify the specific conditions under which an overdense region can form a black hole.

In this work, within the peak theory framework, we consider the collapse of initially non-spherical overdense regions -- treated as ellipsoids -- and apply the hoop conjecture to determine the final outcome. 
On one hand, we obtain a more accurate expression for the formation fraction $\beta$.
On the other hand, we demonstrate that for the vast majority of regions that eventually collapse into PBHs, the tidal torques are extremely small, implying $\langle \tilde{a} \rangle_{\rm BH} \sim \sigma_h^\frac{1}{2}\ll 1$. 
This conclusion is contrary to the previous studies \cite{Harada:2017fjm,Saito:2024hlj}.
To avoid confusion, we point out that $\langle \tilde{a} \rangle \sim \sigma_h^{-\frac{1}{2}}$ and $\langle \tilde{a} \rangle_{\rm BH} \sim \sigma_h^\frac{1}{2}$ are essentially not in contradiction. The former represents the average spin of all overdense regions that eventually collapse, while the latter is the average for those that actually form black holes.

This paper is organized as follows. 
In section \ref{Preliminary Understanding}, we briefly review the linear perturbation theory in a MD era and introduce the Zel'dovich approximation.
In section \ref{The PBH formation in the Peak Theory}, we employ peak theory together with the Zel'dovich approximation to analyze the evolution of overdense regions, showing that the vast majority of black holes undergo the highly isotropic collapse. Numerical calculations are then used to determine the resulting PBH abundance and mass spectrum.
In section \ref{The PBH spin}, we demonstrate that the PBH spin is very
small when $\sigma_h^*\ll 1$, which is different from the estimates of previous articles, and we present its distribution based on our numerical calculations.
Section \ref{monochromatic} is devoted to the special case of a monochromatic power spectrum: 
for $\sigma_h^*\ll 1$, the PBH mass distribution becomes strictly monochromatic and the spin vanishes exactly, and we find that the resulting abundance is 19 times the previous estimate.
In the Appendix, we summarize the peak theory and provide the detailed steps of our numerical procedure for computing the PBH abundance.
The remaining part of this paper presents our conclusions and discussion.
Throughout this paper, we use geometrized units in which both the speed of light and Newton's gravitational constant are unity, $c=G=1$.

\section{Preliminary Understanding} \label{Preliminary Understanding}

Most of this part comes from the textbook and we represent them just as done in Ref.~\cite{Saito:2024hlj}, for the sake of comparison. 

\subsection{The background and perturbation in a matter-dominated era}
In a MD era, the universe could be depicted by the Newtonian gravity. The pressureless fluid elements satisfy the three equations, which are the continuity equation, the Euler equation, and the Poisson equation, respectively,
\begin{gather}
    \left(\frac{\partial\rho}{\partial t} \right)_{\vec r} + \nabla_{\vec r}\cdot  \left(\rho \frac{d\vec r}{dt} \right) = 0, \\
    \frac{d^2\vec{r}}{dt^2} = -\nabla_{\vec r} \Psi, \\
    \Delta_{\vec r} \Psi = 4 \pi \rho.
\end{gather}
As we know, the universe is homogeneous and isotropic on a large scale so we can divide the physical quantities of fluids into two parts: background and perturbation. We can introduce the Eulerian comoving coordinates ${\vec x}\coloneqq{\vec r}/a(t)$ where $a(t)$ is the scale factor. In the background, the fluid element is static at ${\vec x}$, and the energy density is dependent only on cosmic time. The quantities of the background are easily given, as shown by
\begin{gather}
    \rho_b a^3 = \eta_0 = {\rm Const.}, \label{2.4}\\
    \Psi_b = \frac{2\pi}{3}\rho_b r^2, \label{2.5}\\
    H^2 = \frac{8\pi}{3} \rho_b,
\end{gather}
where $\rho_b$ and $\Psi_b$ are the energy density and gravitational potential of the background, respectively; $H \coloneqq \frac{\dot{a}}{a}$ is the Hubble parameter.
From Eqs. (\ref{2.4}) and (\ref{2.5}), we have
\begin{gather}
    a(t) = a_0 t^{2/3}, \\
    \rho_b(t) = \frac{1}{6\pi t^2},
    \end{gather}
where we set $a(t=0)=0$ and $a_0 = (6\pi \eta_0)^{1/3}$.

By subtracting the background, we could get the perturbation equations. Defining the density contrast $\delta\coloneqq\frac{\rho-\rho_b}{\rho_b}$ and gravitational potential perturbation $\psi\coloneqq\Psi - \Psi_b$, we have
\begin{gather}
    \frac{\partial}{\partial t} {\vec u} + H {\vec u} + \frac{1}{a}({\vec u} \cdot \nabla_{\vec x}){\vec u} = -\frac{1}{a}\nabla_{\vec x} \psi,\\
    \frac{\partial}{\partial t}\delta + \frac{1}{a}[\nabla_{\vec x}\cdot{\vec u}+\nabla_{\vec x}\cdot(\delta{\vec u})]=0, \\
    \Delta_{\vec x}\psi = 4\pi\rho_b a^2 \delta,\label{possion}
\end{gather}
with ${\vec u}\coloneqq \frac{d\vec{r}}{dt}-H\vec{r}=a\dot{\vec x}$ the peculiar velocity of the fluid element. In this article, we only focus on the perturbations to the first order. Then, in the Fourier space, we can solve the above coupled equations to the linear order as
\begin{gather}
    \delta_{\vec k}\coloneqq\int\delta(\vec x)\exp{(-i\vec{k}\cdot\vec{x}})d^3\vec{x}=A_{\vec k}t^{2/3}+B_{\vec k}t^{-1}, \label{sol1}\\
    {\vec u}_{\vec k}\coloneqq \int{\vec u}(\vec x)\exp{(-i\vec{k}\cdot\vec{x}})d^3\vec{x}=a_0 \frac{i{\vec k}}{k^2} \left[\frac{2}{3} A_{\vec k} t^{1/3} - B_{\vec k}t^{-4/3}\right] + {\vec C}_{\vec k}t^{-2/3}, \label{sol2}\\
    \psi_{\vec k}\coloneqq\int\psi(\vec x)\exp{(-i\vec{k}\cdot\vec{x}})d^3\vec{x}=-\frac{2}{3}\frac{a_0^2}{k^2}\left[A_{\vec k}+B_{\vec k}t^{-5/3} \right] \label{sol3},
\end{gather}
where ${\vec k}$ is the comoving wave vector. $A_{\vec k}$ and $B_{\vec k}$ are constants and ${\vec C}_{\vec k}$ is a constant vector satisfying ${\vec C}_{\vec k}\cdot{\vec k}=0$. We only keep the $A_{\vec k}$ modes and ignore the modes that decay over time. It is worth noticing that both the density contrast and peculiar velocity keep growing while $\psi$ remains constant over time and is determined by the initial perturbation specified by the power spectrum in \eqref{power}. These behaviors help us to understand the subsequent analysis of the collapse of overdensity. 

\subsection{The Lagrangian coordinates and Zel'dovich approximation}

The bulk fluid dynamics is described by the Eulerian approach as above. However, if one wants to track the evolution of the fluid elements, which is necessary here, instead the Lagrangian approach provides a better option. The Lagrangian comoving coordinate ${\vec q}$, defined as the initial Eulerian comoving coordinate ${\vec x}(t=t_0)$, is simply used to label fluid elements; by definition, it does not depend on time. The motion of the ${\vec q}$-element is described by ${\vec D}(t,{\vec  q})$, which is the displacement vector from the background position $\vec q$ to the Eulerian comoving coordinate of this element at time $t$:
\begin{align}
    {\vec x}({\vec q},t) = {\vec q} + {\vec D}(t,{\vec  q}).
\end{align}
Therefore, the volume elements in two coordinates are related by $d^3 {\vec q} =J d^3 {\vec x} $, where the Jacobi $J({\vec q},t)=({\rm det}( \mathbb{1}_{ij}+\frac{\partial D_i}{\partial q_j}))^{-1}$ evolves and at time $t_0$ it is the unit matrix.

The motion of the fluid element is subject to the conservation of mass of this element, $\rho(t_0) a^3(t_0) d^3 {\vec q}$, which does not change as it moves to the position at $\vec x(\vec q,t)$, thus  
\begin{align}
  \rho(t_0) a^3(t_0) d^3 {\vec q}=\rho(t) a^3(t) d^3 {\vec x} \implies \rho_b(t_0) (1+\delta(t_0)) a^3(t_0)d^3 {\vec q}= \rho_b(t) (1 + \delta(t)) a^3(t) d^3 {\vec x}. 
\end{align}
and therefore, by neglecting $\delta(t_0,\vec q)$ at the initial time, and moreover using $d^3 {\vec q} =J d^3 {\vec x} $ and $\rho_b(t) a^3(t) = \rho_b(t_0) a^3(t_0) = \eta_0$, one establishes the relation between $\delta$ and ${\vec D}$, 
\begin{align}
    \delta(t,{\vec  q}) = \frac{1}{{\rm det}\left( \mathbb{1}_{ij}+\frac{\partial D_i}{\partial q_j}\right)}-1 \simeq -\nabla_{\vec q} \cdot  {\vec D}(t,{\vec  q}),
\end{align}
where the second equality is accurate to the first order of ${\vec D}$. Substituting it into Eq. \eqref{possion}, we obtain the local relation between the displacement vector and the gravitational potential perturbation
\begin{align}
    \Delta_{\vec q}\psi({\vec q})=-4\pi\rho_b a^2 \nabla_{\vec q} \cdot {\vec D},
\end{align}
which can be solved for growing modes as
\begin{align}\label{D(t)}
    {\vec D}(t,{\vec q}) = -\frac{a(t)}{4\pi\eta_0}\nabla_{\vec q} \psi({\vec q}) \propto a(t).
\end{align}
Therefore, in the linear approximation, the fluid element moves according to the form
\begin{align}
    \vec{x}(\vec{q},t) = \vec{q} -\frac{a(t)}{4\pi\eta_0}\nabla_{\vec q} \psi({\vec q}),
\end{align}
which is referred to as the Zel'dovich approximation.  In this approximation, the motion of the fluid element $\vec q$ is simply determined by the local gradient of the gravitational potential with respect to the Lagrangian coordinates, with the time dependence encoded in $a(t)$.

Let us consider a special case.  In an overdense region near $\vec q=0$, we can perform a Taylor expansion of $\psi$ to the second order of $\vec{q}$,
\begin{align}\label{x:evolve}
    x_i(\vec{q},t) = q_i - \frac{a(t)}{4\pi\eta_0} \frac{\partial \psi}{\partial q_i}(\vec q=0) - \frac{a(t)}{4\pi\eta_0} \frac{\partial^2 \psi}{\partial q_i \partial q_j}(\vec q=0) q_j.
\end{align}
where the last term describes the effect of gravitational tidal force; it's due to the inhomogeneity of the gravitational field and characterizes the relative acceleration of neighboring free falling particles.
Note that the second term on the right side of  Eq. \eqref{x:evolve} does not depend on $\vec q$, thus representing the overall motion of the entire high-density region. If we only focus on its collapse process, this term can be dropped and
therefore
\begin{align}\label{D_exp}
     x_i(\vec{q},t) = q_i+D_i(t, \vec q)\simeq  q_i - \tilde{D}_{ij}(t) q_j,
\end{align}
where
$\tilde{D}_{ij} \coloneqq  \frac{a(t)}{4\pi\eta_0} \frac{\partial^2 \psi}{\partial q_i \partial q_j}(\vec q=0)$ is the gravitational tidal tensor; note that $\tilde{D}_{ij}(t)\propto a(t)$, but it remains small
($\lesssim 1$) until the fluid element drops onto the pancake where $x_i\to 0$. The above equation represents that the collapse of the overdense region is driven by the growing gravitational tidal tensor at $\vec q=0$.

\section{The PBH Formation in the Peak Theory} \label{The PBH formation in the Peak Theory}

\subsection{The initial shape of the overdense region}

PBHs come from the collapse of random overdense regions. We assume that an overdensity encloses a local density maximum $\delta_{\rm peak}$ and the boundary is fixed by $\delta(\vec{q})=0$. 
In the vicinity of the peak (conveniently located at $\vec{q}=0$), the density contrast could be expanded up to the second order in ${\vec q}$, as follows:
\begin{align}\label{delta}
    \delta = \delta_{\rm peak} - \frac{1}{2}\sigma_2 \sum_{i=1,2,3} \lambda_i q_i^2.
 \end{align}
Here $\lambda_i\ (i=1,2,3)$ are the eigenvalues of $-\frac{\delta_{ij}}{\sigma_2}$, with $\delta_{ij}\coloneqq\frac{\partial^2 \delta}{\partial q_i\partial q_j}$ taken at $\vec{q}=0$, and the coordinates $q_i (i=1,2,3)$ are defined in the principal axis coordinate frame of  $-\frac{\delta_{ij}}{\sigma_2}$, which is the frame primarily adopted in this paper.
The parameters $\sigma_j$ with $j=0,1,2$ are defined as the spectral moments of the density perturbation power spectrum $P(k)$, as shown by Eq.~(\ref{power}). Under linear approximation, like the perturbation $\delta$ itself, the spectral moments evolve proportionally to the scale factor: $\sigma_j(t) \propto a(t)$. Hence, we divide $\delta_{ij}$ by $\sigma_2$ to cancel the time dependence. Note that we have chosen the principal axes system of  $-\frac{\delta_{ij}}{\sigma_2}$ to ensure that the mixed second derivatives vanish at the peak, thereby simplifying the expression. As a consequence, we can approximate the boundary as an ellipsoid surface, simply given by 
\begin{align}
\label{ellipsoid surface}
    \frac{q_1^2}{A_1^2} + \frac{q_2^2}{A_2^2} + \frac{q_3^2}{A_3^2}= 1,
\end{align}
where the three time-independent semi-axes of the ellipsoid are given by $A_i^2\coloneqq2\frac{\sigma_0}{\sigma_2} \frac{\nu}{\lambda_i}$ $(i=1,2,3)$ with $\nu\coloneqq {\delta_{\rm peak}}/{\sigma_0}$. The average comoving radius of the ellipsoid is
\begin{align}\label{q0star_Definition}
    q_0 \coloneqq(A_1A_2A_3)^{1/3}=q_0^*\frac{(2\nu)^{1/2}}{(\lambda_1\lambda_2\lambda_3)^{1/6}},
\end{align}
 with $q_0^*\coloneqq \left(\frac{\sigma_0}{\sigma_2}\right)^{1/2}$ the characteristic comoving radius of the power spectrum $P(k)$ (see Eq.  \eqref{k*_def}). 
For later convenience, let us write Eq. (\ref{ellipsoid surface}) in the matrix form $q_i \left(\frac{\Lambda}{2\nu q_0^{*2}}\right)_{ij} q_j=1$, with $\Lambda \coloneqq {\rm diag}(\lambda_1, \lambda_2, \lambda_3)$.

For convenience in the discussion of the collapse process in the next subsection, we introduce the horizon–entry time $t_h$ for the overdense region, defined by the usual condition that its physical scale equals the Hubble horizon:
\begin{align}\label{entry}
    a(t_h)q_0=\frac{1}{H(t_h)}.
\end{align}
Using the MD era scaling $a\propto t^{2/3}$
and $H\propto t^{-1}$, we derive from the above equations that $a(t_h)\propto q_0^2$, indicating that larger ellipsoids enter the horizon at later times. 
During the MD era, the mass $M$ of the overdense region is conserved and hence equals the Hubble mass at the horizon–entry time $t_h$,
\begin{align}\label{mass}
    M = \frac{4\pi}{3} \rho_b(t_h) \frac{1}{ H^3(t_h)} = \frac{1}{2H(t_h)} = \frac{1}{2} a(t_h)q_0= \frac{1}{2}  a(t_h) q_0^* \cfrac{(2\nu)^{1/2}}{(\lambda_1\lambda_2\lambda_3)^{1/6}} \propto q_0^3,
\end{align}
where the second equality follows from the Friedmann equation $H^2=\frac{8\pi}{3}\rho_b$ and the final equality is a result of the ellipse with Eq.  \eqref{q0star_Definition}.

\subsection{The pancake collapse of the ellipsoid}

To analyze the dynamics of the ellipsoid collapse, we consider the overdensity in the vicinity of the peak, described by Eq. (\ref{delta}). 
In the Zel'dovich approximation \eqref{D_exp}, the evolution of the ellipsoid follows
\begin{align}\label{Eu}
    r_i(t) = a(t) x_i(t)  = a(t) T_{ij}(t) q_j,
\end{align}
where $T$ is the time-dependent symmetric evolution matrix defined by 
\begin{align}
    T_{ij}(t) \coloneqq \mathbb{1}_{ij} - \sigma_0(t) \frac{\tilde{D}_{ij}}{\sigma_0}.
\end{align}
Here we have extracted out a factor $\sigma_0(t) \propto a(t)$, which serves as a cosmic time indicator, so that $\frac{\tilde{D}_{ij}}{\sigma_0}$ is time-independent (one may dub it the rescaled tidal tensor).

In general, the principal axis coordinate system of $-\frac{\delta_{ij}}{\sigma_2}$ is not the principal axes of the rescaled cosmic tidal tensor $\frac{\tilde{D}_{ij}}{\sigma_0}$.
Accordingly, one may parameterize it as (see Eq. \eqref{rescaled})
\begin{align} 
\label{D_matrix}
\frac{\tilde{D}_{ij}}{\sigma_0} = 
\begin{bmatrix}
\frac{1}{3}\tilde{D}_A+\tilde{D}_B+\frac{1}{3}\tilde{D}_C  & -w_3 & -w_2 \\
-w_3 & \frac{1}{3}\tilde{D}_A-\frac{2}{3}\tilde{D}_C & -w_1 \\
-w_2 & -w_1 & \frac{1}{3}\tilde{D}_A-\tilde{D}_B+\frac{1}{3}\tilde{D}_C 
\end{bmatrix}.
\end{align}
Here, $\tilde{D}_A$ represents the isotropic part of the collapse, while $\tilde{D}_B$ and $\tilde{D}_C$ denote the anisotropy in the collapse velocity along the principal axes of the ellipsoid. The quantities $w_i\ (i=1,2,3)$ represent the tidal torques, which, in fact, rotate the axes of the ellipsoid and generate angular momentum. Unless one considers the monochromatic power spectrum ($\gamma \rightarrow 1$), in which $w_i \lesssim  O\left(\sqrt{1-\gamma^2}\right) \rightarrow 0$, the tidal torque acting on the ellipsoid generally cannot be neglected, and consequently, the overdense region was expected to acquire a considerable spin during the collapse process. 
(See Eq.~\eqref{gamma_definition} and the subsequent text for the definition and properties of $\gamma$.)
We will turn back to this point in subsection  \ref{Black hole formation criterion}. 

Since Lagrangian comoving coordinates are time-independent, the ellipsoid's boundary consistently satisfies Eq.~\eqref{ellipsoid surface}. That is to say, after using Eq.~(\ref{Eu}), it is straightforward to derive that the boundary of the overdensity follows the equation
\begin{align}\label{QuartFM}
    r_i \left(\frac{T(t)^{-1}\Lambda T(t)^{-1}}{2\nu a^2(t) q_0^{*2} }\right)_{ij} r_j=1,
\end{align}
and therefore it remains an evolving ellipsoidal surface in the Eulerian coordinate system. 

To elucidate the picture of collapse, we denote the three eigenvalues of the rescaled cosmic tidal tensor $\frac{\tilde{D}_{ij}}{\sigma_0}$ as $d_i$ ($i=1,2,3$).
So, in its principal axis frame, the evolution equations take the form of
\begin{equation}\label{evolve}
    \begin{cases}
        r_1(t) = a(t) (1 - \sigma_0(t) d_1) \tilde q_1,\\
        r_2(t) = a(t) (1 - \sigma_0(t) d_2) \tilde q_2,\\
        r_3(t) = a(t) (1 - \sigma_0(t) d_3) \tilde q_3.
    \end{cases}
\end{equation}
Here, $r_i$ and $\tilde q_i$ denote, respectively, the Eulerian physical coordinate and the Lagrangian comoving coordinate diagonalizing $\frac{\tilde{D}_{ij}}{\sigma_0}$ of an arbitrary point within the ellipsoid.
From Eq. \eqref{evolve} one can see that, along the directions with $d_i>0$, one has $r_i \propto a(t)$ when $\sigma_0(t)$ is small; that is to say, the ellipsoid initially expands with the cosmic background. After $\sigma_0(t)$ grows to $\frac{1}{2d_i}$, when $\dot{r}_i=0$, the ellipsoid starts to contract along this axis. By contrast, along the directions with $d_i\leq 0$, the ellipsoid continually expands. Let $d_{\rm max}\coloneqq \max_i d_i$. 
Then, at time $t_c$ when $\sigma_0(t_c)= {1}/{d_{\rm max}}$, the ellipsoid collapses into a two-dimensional ellipse. We denote this $t_c$ as the collapse time. This process is also referred to as "pancake collapse".

Furthermore, the resulting ellipse can be determined by the eigenvalues of $T(t_c)\Lambda^{-1} T(t_c)$; this matrix encodes the geometric information of the quadratic form matrix in Eq.~\eqref{QuartFM} at the collapse time $t_c$. Denote its positive eigenvalues as $x_l^2$ and $x_s^2$ ($x_l>x_s>0$, and these two notational labels are unrelated to the Eulerian comoving coordinate $x_i$), then the semi-major and semi-minor axes are $r_l\coloneqq (2\nu)^{1/2} a(t_c) q_0^*x_l $ and $r_s \coloneqq(2\nu)^{1/2} a(t_c) q_0^*x_s$, respectively.
The circumference $C$ of this ellipse is given by $C = 4 r_l  E(e)$,
where $e = \sqrt{1 -  {x_s^2}/{x_l^2}}$ is the eccentricity and $E(e)$ is the complete elliptic integral of the second kind.

\subsection{Black hole formation criterion } \label{Black hole formation criterion}
Does the pancake collapse eventually form a PBH?
The hoop conjecture \cite{Klauder:1972je,Misner:1973prb} furnishes a criterion. It states that black holes with horizons form if and only if a mass $M$ is compacted into a region whose circumference $C$ in every direction is approximately less than $4\pi M$. For the pancake, a black hole can form if and only if $C\lesssim 4\pi M $.
Here the pancake's mass $M$ is determined by Eq. (\ref{mass}) and its circumference is given as $C = 4 r_l  E(e) = 4 (2\nu)^{1/2} a(t_c) q_0^*x_l E(e)$. Now, we can write down the condition for PBH formation as
\begin{align}\label{thor}
    x_l \lesssim x_{\rm th} \coloneqq \frac{\pi d_{\rm max}}{2E(e) (\lambda_1 \lambda_2 \lambda_3)^{1/6}} \sigma_h,
\end{align}
where we have used the relations $ {a(t_h)}/{a(t_c)}= {\sigma_0(t_h)}/{\sigma_0(t_c)} = d _{\rm max}\sigma_h$ with $\sigma_h \coloneqq \sigma_0(t_h)$, the variance of density contrast at horizon cross. Note that $\sigma_h$ is related to the average comoving radius of the overdense region, $q_0$, i.e. $\sigma_h=\sigma_0(t_h)\propto a(t_h(q_0)) \propto q_0^2 \propto M^ {{2}/{3}}$, and thus one may denote it as $\sigma_h(q_0)$, explicitly given by
\begin{align}
    \sigma_h(q_0) = \left(\frac{q_0}{q_0^*} \right)^2\sigma_h^* = \frac{2\nu}{(\lambda_1\lambda_2\lambda_3)^{1/3}} \sigma_h^*,
\end{align}
where $\sigma_h^* =\sigma_h(q_0^*)$ is determined by the given power spectrum $P(k)$,
and characterizes the fluctuation size $\sigma_h$. The density fluctuations are usually very small, so we have $\sigma_h^*\ll 1$. Substituting the above expression into Ineq. \eqref{thor}, $x_{\rm th}$, the threshold for black hole formation, is simply expressed in terms of the initial ellipsoid parameters associated with the given peak and the overall perturbation power spectrum: 
\begin{align}
\label{condition}
x_l \lesssim x_{\rm th} =\frac{\pi \nu d_{\rm max}}{ E(e)(\lambda_1 \lambda_2 \lambda_3)^{1/2}} \sigma_h^*.
\end{align}
Then, based on this condition, it is convenient to determine which peaks can collapse into black holes. We discuss the consequence of this condition in the following. 

In this work, a generic peak can be labeled by nine parameters $\left(\nu, {\vec \lambda}, {\vec w}, \tilde{D}_B, \tilde{D}_c \right)$ with $\nu=\tilde{D}_A$. 
From the peak number density distribution \eqref{peak_distribution}, it is known that the vast majority of peaks satisfy $\left(\nu, {\vec \lambda}, {\vec w}, \tilde{D}_B, \tilde{D}_C \right) \lesssim \mathcal{O}(1)$.

Then $d_{\rm max}$ is also $\sim \mathcal{O}(1)$, since it is nothing but the largest eigenvalue of the matrix $\frac{\tilde{D}_{ij}}{\sigma_0}$, whose elements are given in Eq. (\ref{D_matrix}), consisting of ${\vec w}$ and $\tilde{D}_{A}$, etc. With such estimations, it is seen that the PBH formation condition (\ref{condition}) actually implies the smallness of $ x_l \lesssim \mathcal{O}(\sigma_h^*)$. Further recalling that $x_l^2$ is the eigenvalue of $T(t_c)\Lambda^{-1}T(t_c)$, where $\Lambda \sim \mathcal{O}(1)$, 
we must require ${\rm abs}[T_{ij}(t_c)]={\rm abs}\left[\mathbb{1}_{ij}-\frac{1}{d_{\rm max}}\frac{\tilde{D}_{ij}}{\sigma_0}\right]\lesssim \mathcal{O}(\sigma_h^*)$. In turn, we must have $\frac{\tilde{D}_{ij}}{\sigma_0} = d_{\rm max} \mathbb{1}_{ij} + \mathcal{O}(\sigma_h^*)$, namely the hierarchical structure of the elements
\begin{align}\label{PBH:con}
    &\begin{cases}
        \tilde{D}_B, \tilde{D}_C, {\vec w} \lesssim  \mathcal{O}(\sigma_h^*), \\
        d_{\rm max} = \frac{1}{3}\tilde{D}_A  +  \mathcal{O}(\sigma_h^*)=\frac{1}{3}\nu  +  \mathcal{O}(\sigma_h^*),
    \end{cases}
\end{align}
which are verified to be correct in our numerical calculations.

Several important comments are in order based on Eq. \eqref{PBH:con}:
\begin{itemize}
    \item First, for the overdense regions that eventually collapse into PBHs, when $\sigma_h^*\ll 1$, the vast majority undergo highly isotropic collapse, in agreement with the well-known treatment in in Refs. \cite{Khlopov:1980mg,Harada:2016mhb}. 
    \item Second, this implies that the spins of PBHs generated by tidal torques ($w_i \lesssim  \mathcal{O}(\sigma_h^*)$) are exceedingly small in the regime of  $\sigma_h^*\ll 1$. This result is the most important conclusion in this work, which is contrary to the previous estimation in Refs ~\cite{Harada:2017fjm,Saito:2024hlj}, and we will discuss this process in detail in section  \ref{The PBH spin}. 
    \item  Finally, incidentally, the above discussion does not impose any additional restrictions on the parameter $ \vec \lambda$; therefore, the initial shape of the overdense region can significantly deviate from sphericity. Consequently, this means that it is unnecessary to treat the overdense region as spherical as in Ref. \cite{Harada:2016mhb} if we use peak theory.
\end{itemize}

It is worth noting that for PBHs formed from the overdense region with $\nu \gg 1$, their typical dimensionless spin is also very small. 
We discuss this case in detail in subsection \ref{large_nu}, where a hierarchical structure distinct from that of Eq. \eqref{PBH:con} is exhibited.

\subsection{The PBH abundance and mass function} \label{pbh abundance and mass function}

We have successfully filtered out those peaks that can eventually collapse into black holes by imposing the condition \eqref{condition} or \eqref{PBH:con}. 
Based on the peak number density distribution \eqref{peak_distribution}, we can then compute the abundance $\beta$ of PBHs, defined as the fraction of the total energy of the Universe.

For convenience, let us consider a spherical region with a comoving radius $q_0^*$, the characteristic comoving radius of the power spectrum. Within this sphere, the number of peaks with parameters ranging from $\nu$ to $\nu+d\nu$, $\lambda_i$ to $\lambda_i+d\lambda_i$, $w_i$ to $w_i+dw_i$, $\tilde{D}_B$ to $\tilde{D}_B+d\tilde{D}_B$, and $\tilde{D}_c$ to $\tilde{D}_c+d\tilde{D}_c$ is given by
\begin{align}
   &\frac{4\pi q_0^{*3}}{3} n_{\rm peak}\left(\nu, \vec{\lambda}, \vec{w}, \tilde{D}_B, \tilde{D}_c \right)  d\nu  d^3\lambda_i d^3w_i d\tilde{D}_B d\tilde{D}_c \notag \\
   =& \frac{4\pi A \gamma^{-3/2}}{3} \exp \left(-\frac{1}{2}Q_3\right) \lambda_1 \lambda_2 \lambda_3 (\lambda_2-\lambda_3) (\lambda_1-\lambda_3) (\lambda_1-\lambda_2)
    d\nu  d^3\lambda_i d^3w_i d\tilde{D}_B d\tilde{D}_c,
\end{align}
where we have employed the definitions of the comoving radius of the ellipsoid $q_0$ 
and the relative width of the power spectrum $\gamma$,  (see Eqs.~\eqref{q0star_Definition} and \eqref{gamma_definition}).
Since only those peaks satisfying $x_l \lesssim x_{\rm th}$ can form black holes, the total number of PBHs within the sphere is
\begin{align}
\label{nbh_int}
    N_{\rm BH} = \frac{4\pi A \gamma^{-3/2}}{3} \int &\exp \left(-\frac{1}{2}Q_3 \right) \lambda_1 \lambda_2 \lambda_3 (\lambda_2-\lambda_3) (\lambda_1-\lambda_3) (\lambda_1-\lambda_2) \notag \\ 
    &\times \Theta \left(x_{\rm th} - x_l \right) d\nu  d^3\lambda_i d^3w_i d\tilde{D}_B d\tilde{D}_c,
\end{align}
with each black hole having a mass $M$ given by Eq.~\eqref{mass} expressed as
\begin{align}
    M = \left(\frac{q_0}{q_0^*}\right)^3M^*
    = \frac{(2\nu)^{3/2}}{(\lambda_1\lambda_2\lambda_3)^{1/2}}M^*,
\end{align}
with $M^*$ the mass contained inside the comoving sphere of $q_0^*$. Thus, the mass fraction of PBHs within the sphere is given by
\begin{align}
\label{beta_int}
    \beta = \frac{4\pi A \gamma^{-3/2}}{3} \int &\frac{(2\nu)^{3/2}}{(\lambda_1\lambda_2\lambda_3)^{1/2}}\exp \left(-\frac{1}{2}Q_3 \right) \lambda_1 \lambda_2 \lambda_3 (\lambda_2-\lambda_3) (\lambda_1-\lambda_3) (\lambda_1-\lambda_2) \notag \\ 
    &\times \Theta \left(x_{\rm th} - x_l \right) d\nu  d^3\lambda_i d^3w_i d\tilde{D}_B d\tilde{D}_c. 
\end{align}
Since the peak distribution is homogeneous throughout the Universe, this also represents the overall abundance of black holes.
This nine-fold integral involves the step function $\Theta \left(x_{\rm th} - x_l \right)$ and is thus very complex, requiring numerical methods detailed in Appendix~\ref{sec:numerics} for its evaluation.

\begin{figure}
    \centering
    \includegraphics[width=0.75\linewidth]{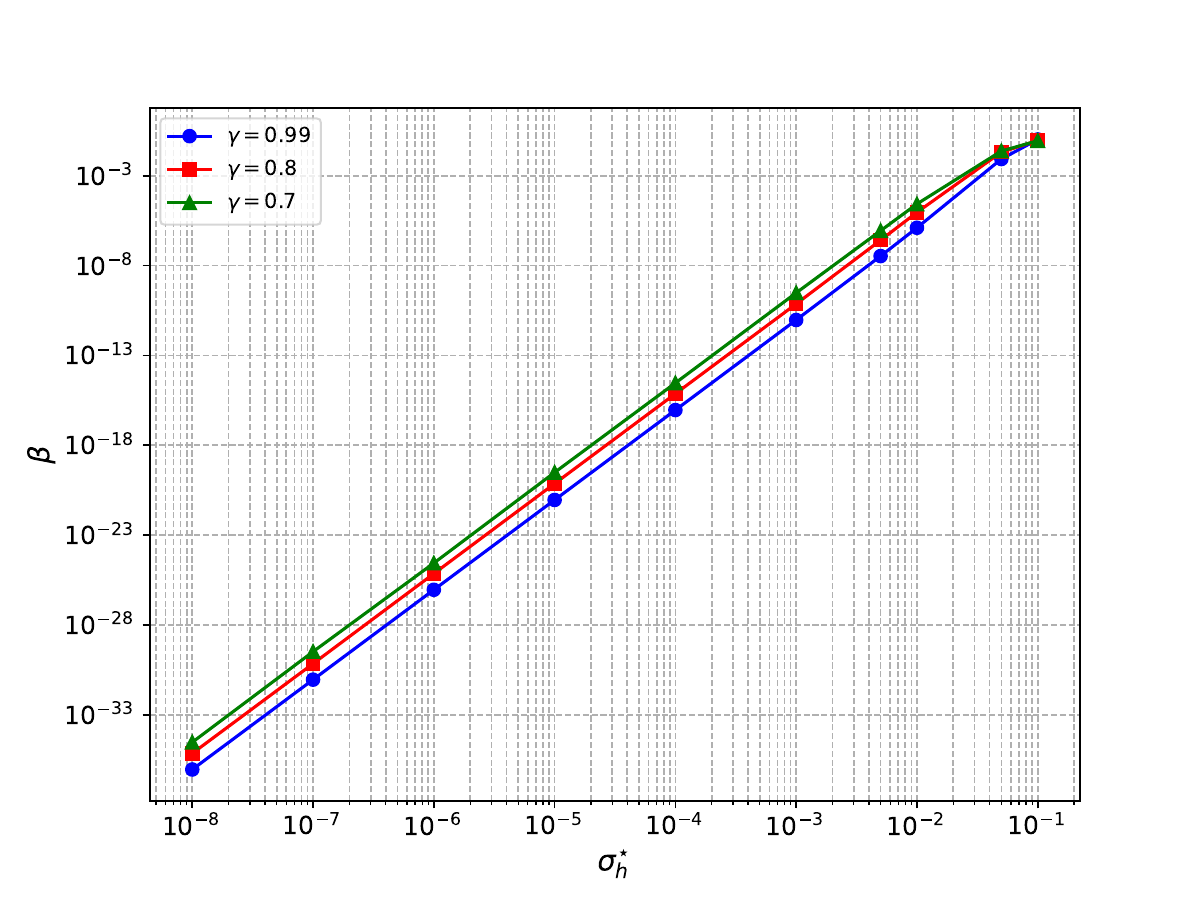}
    \caption{Values of $\beta$ as functions of $\sigma_h^*$ for three relative widths of the power spectrum: $\gamma=0.99 ~(\rm blue),0.8~(\rm red)$ and $0.7~(\rm green)$. In the regime $\sigma_h^*\ll 1$, the curves are well approximated by $\beta \simeq A_\gamma \sigma_h^{*5}$.}
    \label{beta}
\end{figure}
The PBH abundance $\beta$ is only a function of two variables, $\gamma$ and $\sigma_h^*$, which are in turn determined by the power spectrum. In Fig. \ref{beta}, we demonstrate the behavior of $\beta$ varying with $\sigma_h^*$, taking three relative widths of the power spectrum, $\gamma=0.99,0.8$ and $0.7$. It is evident that a larger $\sigma_h^*$ leads to a higher black hole production rate; when $\sigma_h^*\ll 1$, the curve can be approximately fitted by $\beta \simeq A_\gamma \sigma_h^{*5}$.
In Table \ref{beta/sigma_star**5}, we list the values of $\beta / \sigma_h^{*5}$. Approximately, we have $A_{0.99} \simeq 9\times10^3$, $A_{0.8} \simeq 7\times10^4$, $A_{0.7} \simeq 3\times10^5$. 
\begin{table}
    \centering
    \begin{tabular}{ccccccccccc}
        $\sigma^*$& $10^{-8}$&  $10^{-7}$& $10^{-6}$&  $10^{-5}$&  $10^{-4}$&  $10^{-3}$&  0.005&  0.01&  0.05& 0.1\\
        $\gamma=0.99$&  9038&  9020&  9037&  9024&  9070&  9330&  10757&  12647&  27071& 10318\\
        $\gamma=0.8$&  71246&  69994&  71817&  73309&  73584&  73204&  82205&  91338&  65188& 9080\\
        $\gamma=0.7$&  296740&  325299&  276800&  305845&  290865&  300677&  276614&  266296&  75366& 8815\\
    \end{tabular}
    \caption{Values of $\beta/\sigma_h^{*5}$ for $\gamma=0.99,0.8$ and $0.7$.}
    \label{beta/sigma_star**5}
\end{table}

In section \ref{monochromatic}, we instead adopt an alternative characteristic scale $\bar{\sigma}_h$ for  $\sigma_h$. Under this choice one obtains $\beta \simeq 1.06 \bar{\sigma}_h^5$ for $\gamma = 0.99$, which evidently differs by orders of magnitude from $A_{0.99}$.
In fact, it should be emphasized that using either the coefficient $A_\gamma$ or the ratio $\beta/\bar{\sigma}_h^5$ to compare abundances across different values of $\gamma$ is not very meaningful.
This is because $\sigma_h^*$ and $\bar{\sigma}_h$ merely reflect two alternative prescriptions for the "characteristic" value of $\sigma_h$. Different choices of the characteristic scale for $\sigma_h$ can lead to changes of several orders of magnitude in the corresponding coefficient.

\begin{figure}[H]
  \centering            
  \subfloat[] 
  {
      \label{}\includegraphics[width=0.48\textwidth]{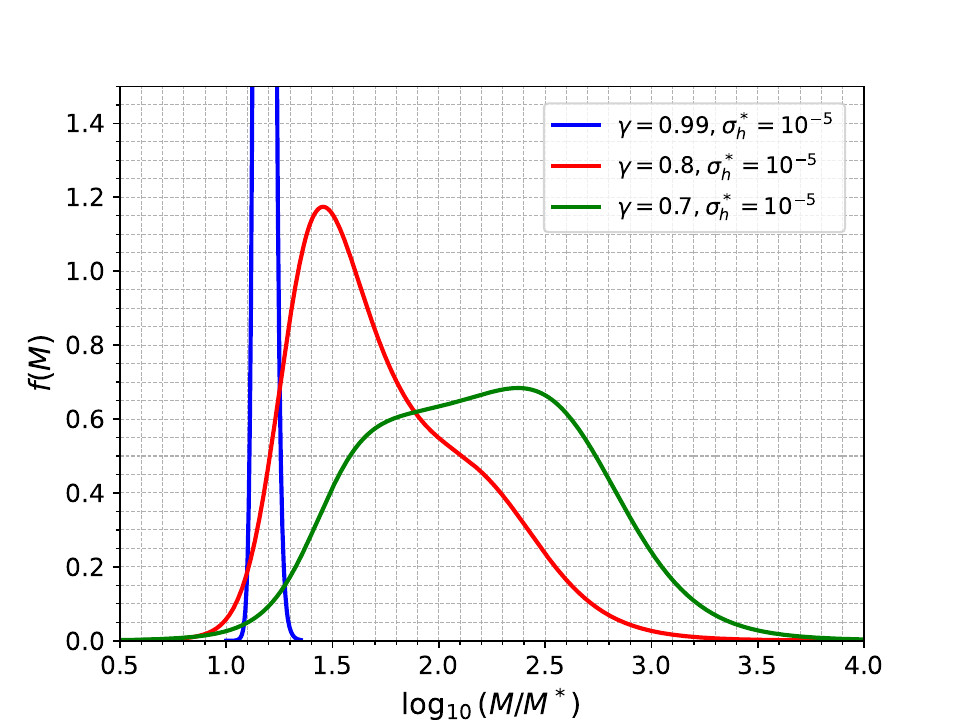}   
  }
  \subfloat[]
  {
      \label{}\includegraphics[width=0.48\textwidth]{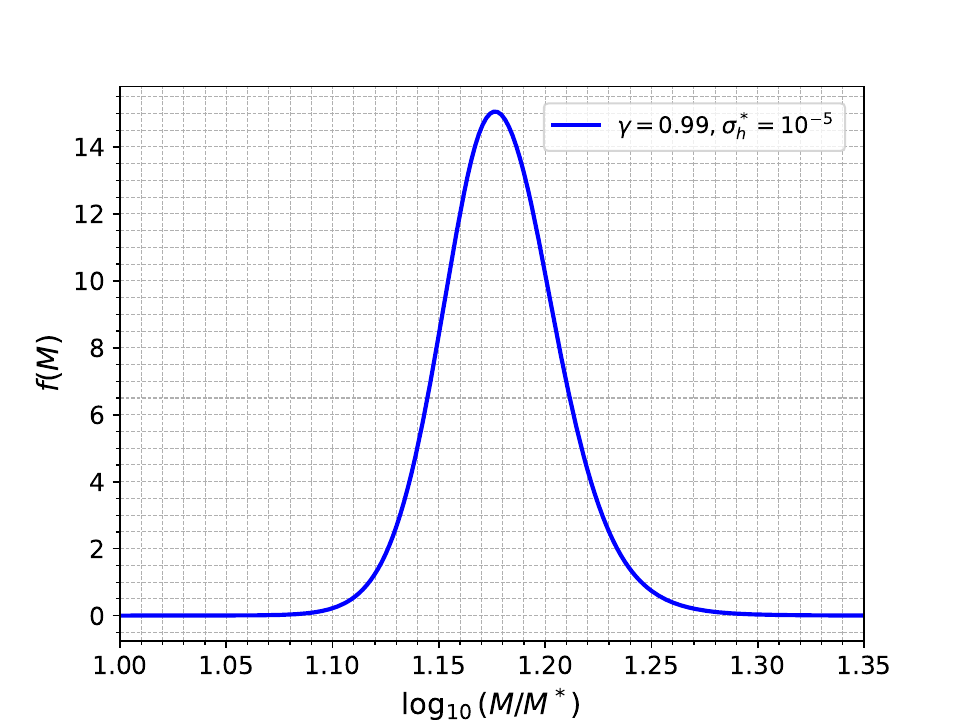}
  }
  \caption{The PBH mass function $f(M)$ at $\sigma_h^*=10^{-5}$ for $\gamma=0.99 ~(\rm blue),0.8~(\rm red)$ and $0.7~(\rm green)$.} 
  \label{mass_function}        
\end{figure}

To quantify the fraction of black holes at different masses, we define the PBH mass function as
\begin{align}
    f(M)\coloneqq \frac{1}{\beta} \frac{d\beta}{d\log_{10} M}.
\end{align}
For $\sigma_h^*=10^{-5}$, we present the mass functions corresponding to $\gamma=0.99, 0.8$ and $0.7$ in Fig. \ref{mass_function}. 

It is seen that, with the increasing width of the power spectrum, the mass distribution of PBHs becomes broader, which is well consistent with the naive expectation. In the case $\gamma = 0.99$, the PBH mass distribution is quite narrow. In section \ref{monochromatic}, we will see that, when $\sigma_h \ll 1$, the mass distribution corresponding to a monochromatic power spectrum is also monochromatic. 

We define the average PBH mass as follows:
\begin{align}\label{PBH:av}
    \bar{M}\coloneqq \frac{\beta}{N_{\rm BH}} M^*. 
\end{align}
Its values under different choices of $\gamma$ and $\sigma_h^*$ are shown in Table \ref{mass_average}. When $\sigma_h^* \ll 1$, our numerical results show that as $\gamma$ decreases, the ratio $\bar{M}/M^*$ increases: for $\gamma=(0.99, 0.8, 0.7)$, one has $\bar{M} \approx (15.1, 40, 100)M^*$.
\begin{table}
    \centering
    \begin{tabular}{ccccccccccc}
        $\sigma_h^*$& $10^{-8}$&  $10^{-7}$& $10^{-6}$&  $10^{-5}$&  $10^{-4}$&  $10^{-3}$&  0.005&  0.01&  0.05& 0.1\\
        $\gamma=0.99$&  15.1&  15.1&  15.1&  15.1&  15.1&  15.1&  15.1&  15.1&  15.9& 18.5\\
        $\gamma=0.8$&  42.9&  42.4&  43.3&  44.0&  43.4&  42.3&  39.7&  36.3&  26.4& 21.3\\
        $\gamma=0.7$&  102.4&  113.9&  95.6&  103.7&  98.7&  98.2&  70.7&  57.7&  29.7& 22.2\\
    \end{tabular}
    \caption{Values of $\bar{M}/M^*$ for $\gamma=0.99,0.8$ and $0.7$.}
    \label{mass_average}
\end{table}

\section{The PBH Spin: Not Large!} \label{The PBH spin}

We will consider the spin up to the first order perturbation. For an ellipsoid around some given peak, the angular momentum of the system in Newtonian mechanics is given by
\begin{align}
    {\vec S} &= \int a^3 d^3{\vec q} \rho_b {\vec r} \times \frac{d{\vec r}}{dt} = \int a^5 d^3{\vec q} \rho_b  {\vec x}\times \left(H{\vec x}+ \dot{{\vec x}} \right) = \eta_0 a^2\int d^3{\vec q} \left({\vec q}+{\vec D}\right) \times \dot{{\vec D}} \notag \\
    &\simeq \eta_0 a^2\int d^3{\vec q} \left({\vec q}\times \dot{{\vec D}} \right) \simeq \frac{2}{3t}\eta_0 a^2\int d^3{\vec q} \left({\vec q}\times {\vec D} \right).   
\end{align}
In the first line, the last equality just states that only the peculiar velocity $\vec u=a \dot{{\vec x}}=a\dot{{\vec D}} $ contributes to the angular momentum; in the second line, we have used Eq.~\eqref{D(t)} and $\dot{\vec D}(t)=H \vec D(t)$ to get the final expression. We could simplify the above expression by expanding ${\vec D}$ around the peak, shown by Eq.~(\ref{D_exp}), to obtain an illustrative expression of the evolving angular momentum of the collapsing region
\begin{align}
\label{angular momentum}
    S_i(t) &\simeq  -\frac{2}{3t}\eta_0 a^2 \epsilon_{ijk}\int d^3{\vec q} \tilde{D}_{kl} q_jq_l = \frac{2}{3t}\eta_0 a^2 \epsilon_{ijk} \tilde{D}_{jl}(t)V_{kl}.
\end{align}
where $V_{kl} \coloneqq \int d^3 {\vec q} q_k q_l$. For an ellipsoid in its principal axis coordinate system, we have
\begin{align}
    V_{kl} = \frac{4\pi}{15} q_0^3
    \cdot {\rm diag} \left(A_1^2, A_2^2, A_3^2 \right).
\end{align}
Introducing the dimensionless spin parameter $\tilde{a}_i \coloneqq  S_i/M^2$, and using the expressions for the angular momentum (Eq.~\eqref{angular momentum}) and mass (Eq.~\eqref{mass}), we obtain
\begin{align}
    \tilde{a}_i(t) = \frac{4}{5} \left(\frac{\sigma_0^3(t)}{\sigma_h^*} \right)^{\frac{1}{2}} \left(\frac{\lambda_1\lambda_2\lambda_3}{8\nu} \right)^{\frac{1}{2}} \epsilon_{ijk} \frac{\tilde{D}_{jl}}{\sigma_0} \Lambda^{-1}_{kl}.
\end{align}
Clearly, the time dependence is encoded in the factor $\sigma_0^{\frac{3}{2}}(t)\propto t$, while others are determined by the peak parameters. In particular, one sees that the spin is sourced by the off-diagonal components of $\frac{\tilde{ D}_{jl}}{\sigma_0}$, i.e. the tidal torque $w_{i}$.

In this work we assume that the final spin of the collapsing region is given by $\tilde{a}_i(t_c)$.
Noting that $\sigma_0(t_c)=1/d_{\rm max}$, we derive
\begin{align}
\label{ac}
    \tilde{a}^2(t_c) &= \frac{2}{25}\frac{\lambda_1\lambda_2\lambda_3}{\nu \sigma_h^* d_{\rm max}^3} \epsilon_{ijk}\epsilon_{imn}\Lambda^{-1}_{kl}\Lambda^{-1}_{ns} \frac{\tilde{D}_{jl}}{\sigma_0} \frac{\tilde{D}_{ms}}{\sigma_0} \notag \\
    & = \frac{2}{25}\frac{\lambda_1\lambda_2\lambda_3}{\nu \sigma_h^* d_{\rm max}^3}
    \left[ 
    w_1^2 \left(\frac{1}{\lambda_2} - \frac{1}{\lambda_3}\right)^2 +
     w_2^2 \left(\frac{1}{\lambda_3} - \frac{1}{\lambda_1}\right)^2 +
      w_3^2 \left(\frac{1}{\lambda_1} - \frac{1}{\lambda_2}\right)^2
    \right],
\end{align}
where $\tilde{a}^2(t_c)\coloneqq  \sum_i \tilde{a}_i^2(t_c)$ denotes the squared spin magnitude.
As we have discussed in subsection~\ref{Black hole formation criterion}, for the vast majority of collapsing regions, all variables in Eq. (\ref{ac}) aside from $\sigma_h^*$ are order-1.
Hence, the mean spin $\langle \tilde{a} \rangle \sim \sigma_h^{*-\frac{1}{2}}$ \footnote{$\langle \tilde{a} \rangle$ and $\langle \tilde{a} \rangle_{\rm BH}$ denote characteristic values of the spin; the exact choice of "mean" does not affect the qualitative discussion presented here. For computational convenience (see Table \ref{spin_average}), we define $\langle \tilde{a} \rangle_{\rm BH}$ simply as the sum of $\tilde{a}$ over all black holes divided by their total number.}, in agreement with
Refs.~\cite{Harada:2017fjm,Saito:2024hlj}. 
In the limit of $\sigma_h^* \ll 1$, this implies $\langle \tilde{a} \rangle \gg 1$.
Refs.~\cite{Harada:2017fjm,Saito:2024hlj} did not, however, incorporate the black hole formation criterion 
from the anisotropic collapse within the peak theory framework as is explicitly noted in Sec.~V and footnote 2 of Ref.~\cite{Saito:2024hlj};
they simply imposed the Kerr bound  $\tilde{a}_{\rm BH}<1$ and hence argued that $\langle \tilde{a} \rangle_{\rm BH} \sim 1$ for $\sigma_h^* \ll 1$.

As we discuss in subsection \ref{Black hole formation criterion}, the hoop conjecture implies that $w_i\lesssim \mathcal{O}(\sigma_h^*)$ for the vast majority of overdense regions that collapse to black holes.
Incorporating this constraint yields $\langle \tilde{a} \rangle_{\rm BH} \sim \sigma_h^{*\frac{1}{2}}$~\footnote{To avoid confusion, we point out that $\langle \tilde{a} \rangle \sim \sigma_h^{*-\frac{1}{2}}$ and $\langle \tilde{a} \rangle_{\rm BH} \sim \sigma_h^{*\frac{1}{2}}$ are essentially not in contradiction. The former represents the average spin of all overdense regions that eventually collapse, while the latter is the average for those that actually form black holes. For the vast majority of overdense regions, they do not form black holes.}, so that the PBH spins are typically very small for $\sigma_h^* \ll 1$.
In Table \ref{spin_average} we list the ratio $\langle \tilde{a} \rangle_{\rm BH}/\sigma_h^{*\frac{1}{2}}$ for $\gamma=0.99, 0.7,0.3$, in agreement with the above estimation. 
\begin{table}
    \centering
    \begin{tabular}{ccccccc}
        $\sigma_h^*$& $10^{-8}$& $10^{-6}$& $10^{-4}$&  $10^{-3}$&  0.005&  0.01\\
        $\gamma=0.99$&  0.0431&  0.0423&  0.0427&  0.0437&  0.0435&  0.0431\\
        $\gamma=0.7$&  1.18& 1.25&  1.33&  1.27&  1.02&  0.91\\
        $\gamma=0.3$&  4.79&  5.44&  6.14&  8.54&  3.38&  2.18\\
    \end{tabular}
    \caption{Values of $\langle \tilde{a} \rangle_{\rm BH}/\sigma_h^{*\frac{1}{2}}$ for $\gamma=0.99,0.7$ and $0.3$.}
    \label{spin_average}
\end{table}
From the table, one sees immediately that:
\begin{itemize}
    \item The smaller the value of $\sigma_h^*$, the smaller the original black hole spin. This has already been discussed in the previous text.
    \item The larger the value of $\gamma$ (the narrower the power spectrum) , the smaller the spin. This behavior follows directly from the peak distribution (Eq. (\ref{peak_distribution})): as $\gamma \rightarrow 1$, the mean tidal torque $w_{i}$ is significantly reduced, and so is the resulting angular momentum. In section \ref{monochromatic} we treat the monochromatic limit, where PBHs are born non‑rotating due to $w_{i}=0$.
\end{itemize}

\begin{figure}[H]
    \centering            
    \includegraphics[width=1\textwidth]{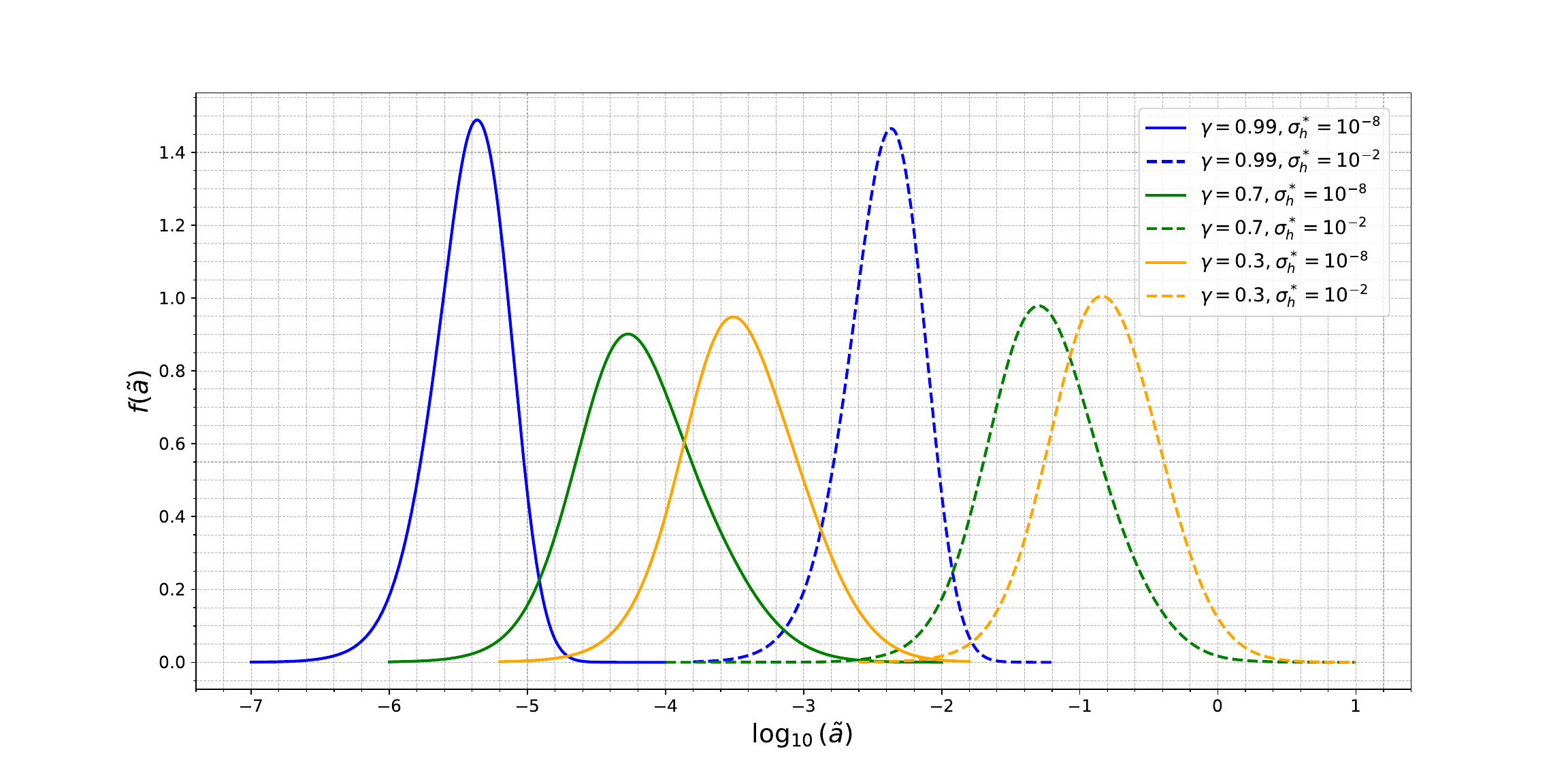}
    \caption{The dimensionless spin distributions of PBHs for $\gamma=0.99$ (blue), $\gamma=0.7$ (green), and $\gamma=0.3$ (orange); solid lines correspond to $\sigma_h^*=10^{-8}$ and dashed lines to $\sigma_h^*=10^{-2}$.} 
    \label{SpinDistribution}  
\end{figure}

We show in Fig. \ref{SpinDistribution} the dimensionless spin distribution of PBHs,
\begin{align}
    f(\tilde{a})\coloneqq \frac{1}{N} \frac{dN}{d \log_{10} (\tilde{a})}.
\end{align}
It is seen that, in agreement with our qualitative analysis, PBH spins are very small for $\sigma_h^*\ll 1$. However, for broader spectra and when $\sigma_h^*$ is not very small (e.g. $\gamma=0.3, \sigma_h^*=10^{-2}$),  the PBH spins can approach 1. Note that the tail of $f(\tilde{a})$ extends beyond $\tilde{a}=1$ for $\gamma=0.3, \sigma_h^*=10^{-2}$, because we have not yet enforced the Kerr bound, which selects only $\tilde{a}<1$. One may apply the same cutoff procedure as in Refs.~\cite{Harada:2017fjm,Saito:2024hlj} to obtain the physical spin distribution of PBHs.

It is noteworthy that Ref. \cite{Saito:2024hlj} argues that the final spin is determined not by the collapse time $t_c$, but by the turnaround time  $t_{\rm ta}$, 
which marks the onset of contraction as the ellipsoid decouples from the expanding background. They posit that angular momentum is conserved from this point onward. In contrast, our study assumes that the evolution of the ellipsoid prior to collapsing to a pancake can be well described by the Zel'dovich approximation, as the perturbation amplitude  $\tilde{D}_{ij}\lesssim 1$ before collapsing to a pancake. This discrepancy arises from the inherent complexities in the nonlinear evolution of perturbations, necessitating various assumptions in approximate treatments. However, the scale factors at the collapse time $a(t_c)$ and at the turnaround time $a(t_{\rm ta})$ are of the same order of magnitude, implying that the spin $\tilde{a}(t_c)$ and $\tilde{a}(t_{\rm ta})$ are also of the same order of magnitude.  Therefore, our conclusions remain unaffected by the choice of assumption regarding the conservation epoch of angular momentum.

\subsection{The PBH spin in the large-$\nu$ limit} \label{large_nu}

As discussed above, for the vast majority of overdensities that collapse to form a PBH, they satisfy the hierarchical structure \eqref{PBH:con} with ($\nu, \vec{\lambda})\lesssim \mathcal{O}(1)$.
Hence we conclude that the typical value of  $\tilde{a}_{\rm BH}$ is extremely small in the regime $\sigma_h^* \ll 1$.
Next, we show that even for peaks with  $\nu \gg 1$, the resulting PBH spin remains small.

Under the high peak approximation in this limit,
\begin{align}
    \lambda_i = \frac{\gamma}{3} \nu + \mathcal{O}(1),
\end{align}
otherwise the peak probability is exponentially suppressed.
This follows from Eq. \eqref{peak_distribution} if one notes that $x^2+(\nu-x\gamma)^2/(1-\gamma^2) = \nu^2 + (x-\nu\gamma)^2/(1-\gamma^2)$).
Since
\begin{align}
    \frac{1}{\lambda_i} - \frac{1}{\lambda_j} = \mathcal{O}(\frac{1}{\nu^2}),
\end{align}
the overdense region is therefore nearly spherical. 
Then, according to Ineq. \eqref{condition}, we estimate
\begin{align}
    x_l \lesssim \mathcal{O}( \nu^\frac{1}{2}\sigma_h^*),
\end{align}
where we have used $\lambda_i \sim \nu$, and $d_{\max}\sim \nu$ (since $d_{\rm max}$  is the largest eigenvalue of $\frac{\tilde{D}_{ij}}{{\sigma_0}}$).
Noting that $x_l^2$ is the eigenvalue of $T(t_c)\Lambda^{-1}T(t_c)$, it follows that
\begin{align}
    {\rm abs}[T_{ij}(t_c)]={\rm abs}\left[\mathbb{1}_{ij}-\frac{1}{d_{\rm max}}\frac{\tilde{D}_{ij}}{\sigma_0}\right]\lesssim \mathcal{O}(\nu \sigma_h^*),
\end{align}
hence
\begin{align}
    &\begin{cases}
        \tilde D_B, \tilde D_C,\vec{w}\lesssim \mathcal{O}(\nu^2 \sigma_h^*), \\
d_{\rm max} =\frac{1}{3}\nu  +  \mathcal{O}(\nu^2 \sigma_h^*).
    \end{cases}
\end{align}

We note that the ratio of anisotropic to isotropic collapse is $\lesssim \mathcal{O}(\nu \sigma_h^*)$, which must be smaller than 1 if the linear perturbation theory is to remain valid.
Using Eq. \eqref{ac}, we then derive $\tilde{a}_{\rm BH} \lesssim \mathcal{O}(\nu^{-\frac{1}{2}} \sigma_h^{*\frac{1}{2}})$ .
Thus the spin acquires an additional suppression factor of $\nu^{-\frac{1}{2}}$, reflecting the high degree of spherical symmetry of the overdensity.

\section{The Monochromatic Power Spectrum for a comparative study} \label{monochromatic}

In this section, we consider the special case $\gamma= 1$ , i.e. $P(k)$ is monochromatic, and compare our results with those in Ref. \cite{Harada:2016mhb}, which did not adopt the peak theory to estimate the PBH abundance. 

In the peak theory, for a monochromatic power spectrum, the tidal forces are no longer independent of the initial ellipsoidal shape. Actually, now the tidal matrix becomes $\frac{\tilde{D}_{ij}}{\sigma_0} = -\frac{\delta_{ij}}{\sigma_2}$, and therefore in the principal axis coordinate system of an ellipsoid, $ \nu=x,\tilde{D}_B = y, \tilde{D}_C = z, \vec{w} = 0$. It means that there are only three independent parameters in the peak theory, chosen to be the eigenvalues of the tidal matrix, $d_i=\lambda_i$ (for $i=1,2,3$) ordered as  $\lambda_1 \geq \lambda_2 \geq \lambda_3$ 
(See Appendix~\ref{appendix:The monochromatic power spectrum} for details). The peak number density per unit comoving volume is then given by  Eq.~(\ref{peak_distribution_monochramatic}). 
In particular, the tidal torque vanishes ($w_i=0$); hence, the ellipsoidal axes do not rotate during collapse and the angular momentum remains zero.

Since $\nu = x = \lambda_1+\lambda_2+\lambda_3$, the initial comoving semi-axes of the overdense ellipsoid are $A_i  = \sqrt{2} q_0^* (\lambda_1+\lambda_2+\lambda_3)^{1/2} / \lambda_i^{1/2}$, and the average radius is $q_0 = \sqrt{2} q_0^* (\lambda_1+\lambda_2+\lambda_3)^{1/2}/(\lambda_1\lambda_2\lambda_3)^{1/6}$, with the corresponding mass given by $M = \frac{1}{2} a(t_h) q_0$. Therefore, at the collapse time $t=t_c$ (with $\sigma_h(t_c)=1/\lambda_1$),  the pancake's two axes are 
\begin{align}
    r_l = a(t_c) \left(1 - \frac{\lambda_3}{\lambda_1} \right)A_3, \\
    r_s= a(t_c) \left(1 - \frac{\lambda_2}{\lambda_1} \right)A_2.
\end{align}
Its circumference is given by $C=4 r_lE(e)$, where $e^2 = 1 - \left( \frac{r_s}{r_l} \right)^2 = 1 - \left( \frac{\lambda_1-\lambda_2}{\lambda_1-\lambda_3} \right)^2 \frac{\lambda_3}{\lambda_2}$.
Invoking the hoop conjecture, the condition for black hole formation is $C \lesssim 4\pi M$, i.e.
\begin{align}
    h(\lambda_1,\lambda_2,\lambda_3) \equiv\frac{2}{\pi} \frac{\lambda_1-\lambda_3}{\lambda_1^2} E(e) \lesssim  
    \frac{\lambda_3^{1/3}}{(\lambda_1\lambda_2)^{1/6}}\sigma_h.
\end{align}
For $\sigma_h \ll 1$ , we have $\lambda_1-\lambda_3 \ll 1$, so the collapse must be highly isotropic; in other words, $\frac{\tilde{D}_{ij}}{\sigma_0} = \lambda_1 \cdot \mathbb{1}_{ij} + \mathcal{O}(\sigma_h^*)$.
This result is entirely consistent with our discussion in subsection  \ref{Black hole formation criterion}.

At this point, we note that for those overdense regions that eventually form black holes, the semi-axes are equal $A_1\simeq A_2\simeq A_3 \simeq  \sqrt{6}q_0^*$.
This implies that only spherical regions with an initial comoving radius $q_0 = \sqrt{6}q_0^*$ can collapse into black holes, corresponding to a mass $M=6\sqrt{6}M^*\approx14.7 M^*$.
By the way, this differs from the high-$\nu$ case we discussed earlier, where even overdense regions that do not collapse into PBHs remain spherically symmetric.
The conclusion shows that when the density perturbation power spectrum  $P(k)$ is monochromatic, the resulting PBH mass distribution is also monochromatic. 
In Table \ref{mass_average}, we present the numerical results for the mass averages at $\gamma=1$; in the limit of $\sigma_h^* \ll 1$, $\bar{M}=14.7M^*$, in agreement with our analytical discussion above. 
For comparison, we also show the results for $\gamma=0.99$, which are very close to those for $\gamma=1$.
\begin{table}
    \centering
    \begin{tabular}{ccccccccccc}
        $\sigma_h^*$& $10^{-8}$&  $10^{-7}$& $10^{-6}$&  $10^{-5}$&  $10^{-4}$&  $10^{-3}$&  0.005&  0.01&  0.05& 0.1\\
        $\gamma=1$& 14.7&  14.7&  14.7&  14.7&  14.7&  14.7&  14.7&  14.7&  15.5& 18.1\\
        $\gamma=0.99$&  15.1&  15.1&  15.1&  15.1&  15.1&  15.1&  15.1&  15.1&  15.9& 18.5\\
    \end{tabular}
    \caption{Values of $\bar{M}/M^*$ for $\gamma=1$ and $0.99$.}
    \label{mass_mono}
\end{table}

Moreover, in the limit $\sigma_h\ll 1$ the black hole formation criterion simplifies to which are very close to those for $h \left(\vec{\lambda}\right)\lesssim \sigma_h$, which is identical to the criterion provided in Ref.~\cite{Harada:2016mhb}.
Next, we discuss the results presented in  Ref.~\cite{Harada:2016mhb}.
In their analysis, they made the following assumptions:
\begin{enumerate}
    \item The initial shape of the overdense region is treated as spherical. 
    \item The collapse process of the overdense region is studied using the Zel'dovich approximation.
    \item The hoop conjecture is adopted as the criterion for black hole formation.
    \item Since the collapse of a spherical region depends solely on the eigenvalues of the rescaled tidal tensor, they used the Doroshkevich probability distribution \cite{1970Ap......6..320D} to compute the black hole abundance $\beta$.
\end{enumerate}

The spherical treatment does not generally hold; for $\gamma \neq 1$ the initial shape of the overdense region can deviate significantly from sphericity.
However, as discussed above, in the case of $\gamma=1$ and $\sigma_h \ll 1$, only spherical regions can form black holes. Thus, we regard their treatment as effectively corresponding to the special case of a monochromatic power spectrum.
Since both our work and Ref. \cite{Harada:2016mhb} adopt the Zel'dovich approximation and the hoop conjecture, the black hole formation condition in the $\sigma_h\ll 1$ limit is the same, namely $h \left(\vec{\lambda} \right)\lesssim \sigma_h$.

However, Ref. \cite{Harada:2016mhb} employs the Doroshkevich probability distribution,
\begin{align}
    \omega \left({\vec \lambda}\right) 
    = & \frac{3375}{8\sqrt{5}\pi}
    \exp \left[ 
    -\frac{1}{2}(\lambda_1 + \lambda_2 + \lambda_3)^2 
    -\frac{5}{4}\left\{ (\lambda_1-\lambda_2)^2+(\lambda_2-\lambda_3)^2+(\lambda_1-\lambda_3)^2 \right\}
    \right] \notag \\
    &\cdot (\lambda_1-\lambda_2)(\lambda_2-\lambda_3)(\lambda_1-\lambda_3),
\end{align}
with $\int d\vec{\lambda} \omega\left({\vec \lambda}\right) =1$, and argues that the black hole abundance is 
\begin{align}
    \beta = \int d\vec{\lambda} \omega\left({\vec \lambda}\right) \Theta\left(\sigma_h-h \left(\vec{\lambda} \right)\right)\simeq 0.056 \sigma_h^5.
\end{align}

In fact, Doroshkevich probability distribution represents the probability distribution of the eigenvalues $\lambda_1, \lambda_2$ and $ \lambda_3$ of the rescaled tidal tensor within a localized overdense region. Therefore, this result actually represents the probability that an arbitrary peak collapses into a black hole, that is, the fraction of peaks forming PBHs. By contrast, the peak theory counts the distribution of peaks throughout the entire space, and therefore it provides a more appropriate way to compute the overall black hole abundance, i.e., the peak number density given by Eq. (\ref{peak_distribution_monochramatic}).

The procedure for calculating the PBH abundance is the same as in Section \ref{pbh abundance and mass function}, so we do not repeat it here. In Table \ref{beta_mono}, we list the values of $\beta/\bar{\sigma}_h^5$ for both $\gamma=1$ and $\gamma=0.99$, where we have introduced the averaged variance corresponding to the averaged PBH mass 
\begin{align}
    \bar{\sigma}_h \coloneqq\sigma_h \left(\bar{M}\right) = \left(\frac{\bar{M}}{M^*} \right)^{2/3} \sigma_h^*.
\end{align}
In the limit $\sigma_h^* \ll 1$, for the monochromatic power spectrum we have $\bar{\sigma}_h = \sigma_h{(M)}= 6\sigma_h^*$; for $\gamma=0.99$, $\bar{\sigma}_h = 6.1\sigma_h^*$.
It can be seen that in the monochromatic case, we have
\begin{align}
    \beta \simeq 1.08 \sigma_h^5,
\end{align}
which is 19 times the result reported in Ref. \cite{Harada:2016mhb}.
For comparison, we find that the results for $\gamma=0.99$ are very close to those for $\gamma=1$.
\begin{table}
    \centering
    \begin{tabular}{ccccccccccc}
        $\sigma_h^*$& $10^{-8}$&  $10^{-7}$& $10^{-6}$&  $10^{-5}$&  $10^{-4}$&  $10^{-3}$&  0.005&  0.01&  0.05& 0.1\\
        $\gamma=1$& 1.08&  1.08&  1.08&  1.08&  1.09&  1.12&  1.28&  1.51&  2.77& 0.67\\
        $\gamma=0.99$&  1.06&  1.07&  1.07&  1.06&  1.06&  1.10&  1.26&  1.49&  2.69& 0.62\\
    \end{tabular}
    \caption{Values of $\beta/\bar{\sigma}_h^5$ for $\gamma=1$ and $0.99$.}
    \label{beta_mono}
\end{table}

\section{Conclusion and Discussion}

In this work, we have presented a detailed analysis of PBHs formed during the MD era within the framework of peak theory.
PBHs originate from the collapse of overdense regions, which we model by assuming each overdensity encloses a local density maximum, $\delta_{\rm peak}$, with its boundary defined by  $\delta(\vec{q})=0$. 
To simplify the collapse dynamics, we approximate each overdense region as an ellipsoid and employ the Zel'dovich approximation to follow its evolution. 
We carefully analyzed the formation conditions of PBHs by the hoop conjecture and found that only regions undergoing highly isotropic collapse can eventually form a black hole. 
We computed the PBH abundance $\beta$ and found that, in the regime $\sigma_h^*\ll 1$, it obeys the scaling law $\beta \simeq A_\gamma \sigma_h^{*5}$.
Moreover, we find that PBHs acquire only a small dimensionless spin, $\langle \tilde{a} \rangle_{\rm BH} \sim \sigma_h^{* \frac{1}{2}} \ll 1$ for $\sigma_h^*\ll 1$, which is contrary to previous results. 
In particular, we consider the special case of a monochromatic power spectrum ($\gamma=1$).
Unlike the broad spectrum, in the regime $\sigma_h^*\ll 1$, only a spherical region of comoving radius $q_0 = \sqrt{6}q_0^*$ can collapse to form a black hole.
Consequently, the mass distribution of PBHs are monochromatic, with $M=6\sqrt{6}M^*$, and their angular momentum vanishes exactly.
Our calculation yields a PBH abundance $\beta \simeq 1.08 \sigma_h^5$, which is 19 times the estimate given in Ref. \cite{Harada:2016mhb}.

Despite our detailed analysis, several important issues remain open for future work. 

\begin{itemize}
\item Our calculation of PBH spins is carried out only to first order in the displacement vector $\vec{D}$, thereby neglecting nonlinear contributions. It remains unclear whether higher-order terms could make a significant impact on the spin.

\item We have not included the effects of accretion. 
Refs. \cite{DeLuca:2021pls, deJong:2021bbo, deJong:2023gsx} have performed exploratory studies of accretion during an MD era and its influence on PBH formation. In general, regions that can not directly collapse into black holes might do so through accretion of the surrounding background matter. Moreover, already formed PBHs can grow by further accretion of ambient material, thereby modifying their mass distribution. Because the accreted matter carries little or no angular momentum, the spin of an existing PBH may be reduced; as a result, PBHs that are born with high spin (i.e., in broad spectra and when $\sigma_h^*$ is not very small) could be spun down to negligible rotation. However, Ref. \cite{Kang:2024trj} have proposed a "short EMD" scenario in which, immediately after PBH formation, the EMD era ends via prompt reheating into the RD era, potentially preventing further spin reduction.

\item It should be emphasized that, when estimating the abundance of primordial black holes (PBHs) from broad power spectra, one must introduce a window function $\tilde{W}(R,k)$ satisfying $\tilde{W}(R,k=0)=1$ and  $\tilde{W}(R,k \gg R^{-1})=0$ in order to smooth out inhomogeneities on scales smaller than $R$ 
with $R$ being the scale of interest (see Ref.~\cite{Yoo:2020dkz} for PBH formation in an RD era). 
This corresponds to the removal of the small-scale components of the power spectrum, i.e.,  $P(k)\rightarrow P(k) \tilde{W}^2(R,k)$.
Otherwise, due to the contamination from small-scale fluctuations, we would be unable to correctly apply peak theory to statistically account for overdensities on the scales of interest.
While the specific choice of window function and smoothing scale does affect quantitative estimates of PBH abundance and spin, these choices do not change the results presented in this paper, since they merely modify the functional form of $P(k)$.
However, Ref.~\cite{Harada:2022xjp} has demonstrated that small-scale perturbations, through their nonlinear growth, generate the velocity dispersion to impede the gravitational collapse of larger-scale modes.
As a result, perturbations on scales $<R$ may impart physical effects that cannot be fully erased by smoothing, leading to a suppressed PBH formation possibility.
Additionally, as Fig. \ref{SpinDistribution} illustrates, a broad power spectrum offers the potential for PBHs with large initial spins during a MD era; yet the velocity dispersion may eliminate this potential by suppressing PBH formation from broad spectra. These details are beyond the scope of this work, and we look forward to future research on the topic.

\item Other factors--such as inhomogeneity \cite{Khlopov:1980mg,polnarev1982dustlike,Kokubu:2018fxy}, have also been identified as playing significant roles in PBH formation. Integrating all of these effects into a unified treatment remains an outstanding challenge.
\end{itemize}

In summary, the formation and evolution of PBHs during an MD era constitute a richly complex problem. We look forward to future investigations that will deepen our understanding.

\appendix
\section{The Peak Theory}

In this section, we briefly summarize aspects of the peak theory relevant to our study, drawing upon Refs. \cite{Bardeen:1985tr,heavens1988tidal}.
The peak theory provides a statistical framework for analyzing the properties of peaks in random fields, which is instrumental in understanding the formation of structures such as PBHs.

We assume that the gravitational potential perturbation $\psi$ is a random Gaussian field. Then, in the case of linear approximation, all perturbative variables can be regarded as random Gaussian fields. In this article, we consider the following sixteen variables, $V_i = \left\{\delta, \delta_i, \delta_{ij}, \tilde{D}_{ij} \right\}$. 
Among them, the first three sets describe the density contrast, up to the gradient $\delta_i \coloneqq \frac{\partial \delta}{\partial q_i}$ and second derivative $\delta_{ij}\coloneqq \frac{\partial^2 \delta}{\partial q_i \partial q_j}$. While the last set, the gravitational tidal tensor $\tilde{D}_{ij}({\vec q_0}) \coloneqq  \frac{a}{4\pi\eta_0} \frac{\partial^2 \psi}{\partial q_i \partial q_j}$, describes the evolution of the density contrast in this region. We will soon see that this set is not independent of the other sets due to the fact that, in the Fourier space, actually both $\delta_{\vec k}$ and $\psi_{\vec k}$ follow the same mode, see Eqs. \eqref{sol1} and \eqref{sol3}. 

In general, the $n$-dimensional joint Gaussian probability of $V_i$ takes the form of
\begin{align}
    f(V_i)d^{n}V_i = \frac{1}{(2\pi)^{n/2}({\rm det}M)^{1/2}}
    \exp\left[ -\frac{1}{2} (V_i-\langle V_i \rangle)M^{-1}_{ij}(V_j-\langle V_j \rangle) \right]d^{n}V_i ,
\end{align}
where $M$ is the symmetric covariance matrix, the core of the peak theory. Concretely, $M_{ij}$ is the correlation function $M_{ij}\left(\vec x,\vec x+\vec r\right)\coloneqq\left\langle \left(V_i(\vec{x})-\langle V_i \rangle\right) \left(V_j(\vec{x}+\vec{r})-\langle V_j \rangle\right) \right\rangle$, reduced to the function of $r$ only for the sake of statistically uniform and isotropic; we are interested in $r=0$ in this work. $M_{ii}$ is the variance of the variable $V_i$, while the non-diagonal element $M_{ij}$ ($i\neq j$) represents the linear correlation between $V_i$ and $V_j$. In this article, all the mean values vanish, $\langle V_i \rangle=\int V_i f(V_i)dV_i=0$. 

To calculate the correlation matrix, we work in the Fourier space, $\delta_{\vec k}\coloneqq\int\delta\left(\vec x\right)\exp\left(-i\vec{k}\cdot\vec{x}\right)d^3\vec{x}$, etc. In cosmology, $P(k)$, the power spectrum of the density contrast $\delta$ field is the ensemble average of the  Fourier component modulus squared, 
\begin{align}\label{power}
    \langle \delta_{{\vec k}}^* \delta_{{\vec k}'} \rangle = (2\pi)^3 \delta^3 \left({\vec k} - {\vec k}'\right) \frac{2\pi^2}{k^3} P(k=|{\vec k}|),
\end{align}
with the delta function indicating that there is no statistical correlation between different modes. 
Then, all correlation functions $M_{ij}$ can be expressed by $P(k)$. 
With Eqs. \eqref{sol1}, \eqref{sol3} and \eqref{power}, the nonzero correlation functions are calculated to be \footnote{
As an example, we present the computation of one of the correlation functions:
\begin{align*}
   \langle\delta_{11}^2(\vec{q}) \rangle 
   &= \int \frac{d^3 \vec{k}}{(2\pi)^3} \frac{d^3 \vec{k^\prime}}
   {(2\pi)^3} k_1^2 k_1^{\prime 2} \langle \delta_{{\vec k}} \delta^*_{{\vec k}'} \rangle
   e^{i(\vec{k}-\vec{k^\prime})\cdot \vec{q}}
   = \int \frac{d^3 \vec{k}}{(2\pi)^3} \frac{d^3 \vec{k^\prime}}
   {(2\pi)^3} k_1^2 k_1^{\prime 2} 
   (2\pi)^3 \delta^3({\vec k} - {\vec k}') \frac{2\pi^2}{k^3} P(k)
   e^{i(\vec{k}-\vec{k^\prime})\cdot \vec{q}}  \\
   &= \int \frac{k^2 \sin{\theta} dk d\theta d\varphi}{(2\pi)^3} (k\sin{\theta}\cos{\varphi})^4 \frac{2\pi^2}{k^3} P(k)
   = \frac{1}{5}\int d\ln{k} k^4 P(k) = \frac{\sigma_2^2}{5}.
\end{align*}
}
\begin{align}\label{cor}
    &\langle\delta^2\rangle 
    = 3\langle\delta\tilde{D}_{11}\rangle
    = 5\langle\tilde{D}_{11}^2\rangle
    = 15\langle\tilde{D}_{11}\tilde{D}_{22}\rangle
    = 15\langle\tilde{D}_{12}^2\rangle
    =\sigma_0^2  \notag \\
    & \langle\delta\delta_{11}\rangle 
    = -\langle\delta_1\delta_1\rangle
    = \frac{5}{3}\langle\delta_{11}\tilde{D}_{11}\rangle
    = 5\langle\delta_{11}\tilde{D}_{22}\rangle
    = 5\langle\delta_{12}\tilde{D}_{12}\rangle
    =-\frac{\sigma_1^2}{3}  \notag \\
    &\langle\delta_{11}^2\rangle 
    = 3\langle\delta_{11}\delta_{22}\rangle
    = 3\langle\delta_{12}^2\rangle
    =\frac{\sigma_2^2}{5}
\end{align}
with obvious generations to other components. 
Here the $k$-order spectral moment $\sigma_j$ is calculated from the power spectrum by
\begin{align}
    \sigma_j^2 \coloneqq\int d\ln k (k^{2})^jP(k).
\end{align}
It's worth noting that $\sigma_j \propto a(t)$ because $\delta\propto a(t)$ under the linear approximation. In particular, for the special case that $P(k)$ is a monochromatic power spectrum, i.e., $\delta_k$ is not vanishing only at certain mode $k^*$ and then $P(k)\propto\delta(k-k^*)$, one has $\sigma_j=(k^*)^j\sigma_0$.
In the context of a general power spectrum, for convenience, we can define a characteristic comoving wave vector as
\begin{align}
\label{k*_def}
    k^*  \coloneqq  \left(\frac{\sigma_2}{\sigma_0}\right)^{\frac{1}{2}},
\end{align}
which corresponds to a characteristic comoving length $q_0^*=1/k^*$. In particular, $\sigma_0$ is the variance of density contrast (also of the diagonal tilde tensor), while $\sigma_{1}$ and $\sigma_{2}$ determine the variances of $\delta_1$ and $\delta_{11,12}$, respectively.

\subsection{Rescaling and linear combination}

For the sake of discussion and numerical analysis, it is more convenient to change these variables as the following
\begin{align}\label{rescaled}
   &x\coloneqq-\frac{1}{\sigma_2}(\delta_{11}+\delta_{22}+\delta_{33}), \quad y\coloneqq-\frac{1}{2\sigma_2}(\delta_{11}-\delta_{33}),  \quad z\coloneqq-\frac{1}{2\sigma_2}(\delta_{11}-2\delta_{22}+\delta_{33}), \notag \\  
    &\nu\coloneqq\frac{\delta}{\sigma_0},  \quad \tilde{\delta}_i\coloneqq\frac{\delta_i}{\sigma_1}, \quad
    \tilde{\delta}_{ij} \coloneqq \frac{\delta_{ij}}{\sigma_2}\quad (i\neq j), \notag \\
    &\tilde{D}_A\coloneqq\frac{1}{\sigma_0}\left(\tilde{D}_{11}+\tilde{D}_{22}+\tilde{D}_{33} \right), \quad \tilde{D}_B\coloneqq\frac{1}{2\sigma_0}\left(\tilde{D}_{11}-\tilde{D}_{33} \right), \quad \tilde{D}_C\coloneqq\frac{1}{2\sigma_0}\left(\tilde{D}_{11}-2\tilde{D}_{22}+\tilde{D}_{33}\right), \notag \\
     &w_1\coloneqq -\frac{\tilde{D}_{23}}{\sigma_0}, \quad
    w_2\coloneqq -\frac{\tilde{D}_{31}}{\sigma_0}, \quad
    w_3\coloneqq -\frac{\tilde{D}_{12}}{\sigma_0}. 
\end{align}
We can find $\tilde{D}_A=\nu$ by Eq. (\ref{possion}), which implies that only 15 variables out of the original 16 are independent. For the Gaussian probability distribution of these new 15 variables, 
i.e. $V_i = \left\{\nu, \tilde{\delta}_{1,2,3}, x, y, z, \tilde{\delta}_{12,23,31},  w_{1,2,3}, \tilde{D}_B, \tilde{D}_c \right\}$ , 
we have
\begin{align}
    f(V_i) d^{15} V_i
= \frac{1}{(2\pi)^{15/2}({\rm det}M)^{1/2}} \exp\left(-\frac{1}{2}Q_1 \right) d^{15} V_i,
\end{align}
where
\begin{align}
    {\rm det}(M) = \frac{(1 - \gamma^2)^6}{5^{10}3^{11}},
\end{align}
and
\begin{align}
    Q_1 = &x^2 + \frac{(\nu-x\gamma)^2}{1-\gamma^2} + 15y^2 + \frac{15(\tilde{D}_B-y\gamma)^2}{1-\gamma^2}+5z^2 + \frac{5(\tilde{D}_C-z\gamma)^2}{1-\gamma^2} 
    +3 (\tilde{\delta}_{1}^2+\tilde{\delta}_{2}^2+\tilde{\delta}_{3}^2)
    \notag \\
    & +\frac{15(w_1-\gamma \tilde{\delta}_{23})^2}{1-\gamma^2}
    +\frac{15(w_2-\gamma \tilde{\delta}_{31})^2}{1-\gamma^2}
    +\frac{15(w_3-\gamma \tilde{\delta}_{12})^2}{1-\gamma^2}
+15(\tilde{\delta}_{23}^2+\tilde{\delta}_{31}^2+\tilde{\delta}_{12}^2).
\end{align}

In Eq. \eqref{rescaled}, the original variables are rescaled by the corresponding variance, which makes them not evolve over time and moreover of the order of unity. Then, the non-zero correlations are
\begin{align}
    & \langle x^2 \rangle = \langle \nu^2 \rangle = 3\langle \tilde{\delta}_1^2 \rangle = 15\langle \tilde{\delta}_{12}^2 \rangle = 15\langle w_3^2 \rangle = 1, \notag \\
    & \langle x\nu \rangle = 5\langle \tilde{D}_c z \rangle = 15\langle \tilde{D}_B y \rangle = 15\langle \tilde{\delta}_{12} w_3 \rangle = \gamma, \notag \\
    & \langle z^2 \rangle = 3\langle y^2 \rangle = \langle \tilde{D}_C^2 \rangle = 3\langle \tilde{D}_B^2 \rangle  = \frac{1}{5},
\end{align}
and so on. Here,
\begin{align}
\label{gamma_definition}
    \gamma \coloneqq \frac{\sigma_1^2}{\sigma_0 \sigma_2},
\end{align}
which takes values in the interval $[0,1]$ and reflects the relative width of the power spectrum  \cite{Bardeen:1985tr}.
A smaller $\gamma$ corresponds to a broader spectrum, 
while $\gamma=1$ indicates that $P(k)$ is a monochromatic spectrum.
$\gamma = 0$ corresponds to $P(k)$ being independent of $k$.

There is one more simplification can be made via the symmetry of the system. For the matrix $-\frac{\delta_{ij}}{\sigma_2}$, its six degrees of freedom can be expressed in terms of three eigenvalues $\lambda_1\geq\lambda_2\geq\lambda_3$ along with three Euler angles, which specify the orientation of the principal axes.   Since $Q_1$ is invariant under rotations of the coordinate system, we can integrate out the Euler angles to obtain the probability density for the remaining 12 variables \cite{heavens1988tidal}
\begin{align}\label{Gaussf}
    &f\left(\nu, \tilde{\delta}_i, \lambda_i, w_i, \tilde{D}_B, \tilde{D}_c \right) d\nu d^3\tilde{\delta}_i  d^3\lambda_i d^3w_i d\tilde{D}_B d\tilde{D}_c \notag \\
    = &A \exp\left(-\frac{1}{2}Q_2\right)  (\lambda_2-\lambda_3) (\lambda_1-\lambda_3) (\lambda_1-\lambda_2) d\nu d^3\tilde{\delta}_i  d^3\lambda_i d^3w_i d\tilde{D}_B d\tilde{D}_c,
\end{align}
where
\begin{align}
    A &=  \frac{3^{11/2}5^5}{2^{13/2}\pi^{11/2}(1-\gamma^2)^3}, \notag \\
   Q_2 &=  x^2 + \frac{(\nu-x\gamma)^2}{1-\gamma^2} + 15y^2 + \frac{15\left(\tilde{D}_B-y\gamma\right)^2}{1-\gamma^2}+5z^2 + \frac{5\left(\tilde{D}_C-z\gamma\right)^2}{1-\gamma^2} 
    +3 \left(\tilde{\delta}_{1}^2+\tilde{\delta}_{2}^2+\tilde{\delta}_{3}^2\right)
    \notag \\
    & + 15\frac{w_1^2+w_2^2+w_3^2}{1-\gamma^2}.
\end{align}
It is worth noting that the definitions of these variables are formally the same as before, but they are defined in the principal axis reference frame of $-\frac{\delta_{ij}}{\sigma_2}$. Specifically, we have $x=\lambda_1+\lambda_2+\lambda_3, y=(\lambda_1-\lambda_3)/2, z=(\lambda_1-2\lambda_2+\lambda_3)/2$.

\subsection{Comoving peak number density}\label{density:comoving}

Typically, we define a local maximum of the density perturbation $\delta$ as a peak, where $\delta_i=0$.
Thus, a peak can be characterized by the 9 parameters  $(\nu, \vec{\lambda}, \vec{w}, \tilde{D}_B, \tilde{D}_c)$. 
We are particularly interested in the number of peaks per unit comoving volume. For an arbitrary peak, which is assumed to be located at $\vec{q}=0$ without loss of generality, we have $\delta_i \approx  \sum_{j} \delta_{ij}q_j$ in its neighborhood. Consequently, the contribution of the peak to the peak number density is
\begin{align}
    \Delta^{(3)}\left(\vec{q}\right) =| \det (\delta_{ij})| \Delta^{(3)}\left(\vec{\delta}\right)
    = \left( \frac{\sigma_2}{\sigma_1}\right)^3 |\lambda_1\lambda_2\lambda_3| \Delta^{(3)}\left(\vec{\tilde{\delta}}\right),
\end{align}
where we temporarily denote the delta function by $\Delta$ to avoid confusion with the symbol $\delta$.
Since a density maximum also requires that the matrix $\delta_{ij}$ is negative definite, we must have $\lambda_i>0$.

Therefore, in the parameter ranges $\nu$ to $\nu+d\nu$, $\lambda_i$ to $\lambda_i+d\lambda_i$, $w_i$ to $w_i+dw_i$, $\tilde{D}_B$ to $\tilde{D}_B+d\tilde{D}_B$, and $\tilde{D}_c$ to $\tilde{D}_c+d\tilde{D}_c$, the number of peaks per unit comoving volume is given by
\begin{align}
\label{peak_distribution}
    &n_{\rm peak}\left(\nu, \vec{\lambda}, \vec{w}, \tilde{D}_B, \tilde{D}_c \right)  d\nu  d^3\lambda_i d^3w_i d\tilde{D}_B d\tilde{D}_c \notag \\
    = & \int d^3 \tilde{\delta}_i
    \left( \frac{\sigma_2}{\sigma_1}\right)^3 |\lambda_1\lambda_2\lambda_3| \Delta^{(3)}\left(\vec{\tilde{\delta}}\right) f\left(\nu, \tilde{\delta}_i, \lambda_i, w_i, \tilde{D}_B, \tilde{D}_c \right)
    d\nu  d^3\lambda_i d^3w_i d\tilde{D}_B d\tilde{D}_c, \notag \\
    = &
   A \left( \frac{\sigma_2}{\sigma_1} \right)^3 \exp \left(-\frac{1}{2}Q_3 \right) \lambda_1 \lambda_2 \lambda_3 (\lambda_2-\lambda_3) (\lambda_1-\lambda_3) (\lambda_1-\lambda_2)
    d\nu  d^3\lambda_i d^3w_i d\tilde{D}_B d\tilde{D}_c,
\end{align}
where
\begin{align}
    \label{Q3}
     Q_3 = x^2 + \frac{(\nu-x\gamma)^2}{1-\gamma^2} + 15y^2 + \frac{15 \left(\tilde{D}_B-y\gamma \right)^2}{1-\gamma^2}+5z^2 + \frac{5 \left(\tilde{D}_C-z\gamma \right)^2}{1-\gamma^2} 
    + 15\frac{w_1^2+w_2^2+w_3^2}{1-\gamma^2}.
\end{align}
$\lambda_i$ are restricted by $\lambda_1\geq\lambda_2\geq\lambda_3>0$ and for the overdensity,  $\nu>0$ are required.

\subsection{The monochromatic power spectrum} \label{appendix:The monochromatic power spectrum}

In the case of a monochromatic power spectrum, $P(k)\propto \delta(k-k^*)$, the density contrast is proportional to the gravitational potential perturbation up to a constant factor, namely
\begin{align}
    \delta(\vec{q},t) = -\frac{3}{2} \left( \frac{k^*}{a_0} \right)^2 t^{2/3} \psi(\vec{q},t)
\end{align}
where we have used Eqs. (\ref{sol1}) and (\ref{sol3}). 
Taking the second derivative with respect to $\vec{q}$ on both sides then yields $\frac{\tilde{D}_{ij}}{\sigma_0} = -\frac{\delta_{ij}}{\sigma_2}$, which implies that, in the principal‐axis coordinate system,
\begin{align} \label{peak_reduced}
    \nu=x,\tilde{D}_B = y, \tilde{D}_C = z, \vec{w} = 0.
\end{align}
Thus, a peak can be described by only three variables; we can choose these to be the eigenvalues $\lambda_i$ of $ -\frac{\delta_{ij}}{\sigma_2}$ (equivalently, of $\frac{\tilde{D}_{ij}}{\sigma_0}$).
In fact, Eq. (\ref{peak_reduced}) follows directly from Eq. (\ref{peak_distribution}): in the limit $\gamma \rightarrow 1$, any peaks not satisfying Eq. (\ref{peak_reduced}) are suppressed to zero measure.
Under a monochromatic spectrum, the number of peaks per unit comoving volume can be obtained in closed form by integrating out $\nu, \tilde{D}_B, \tilde{D}_C$ and $w_i$ from Eq. (\ref{Q3}), giving
\begin{align}
\label{peak_distribution_monochramatic}
       n_{\rm peak}(\lambda_i) d^3\lambda_i
    = &\left(\frac{3}{2}\right)^{7/2} \left(\frac{5}{\pi}\right)^{5/2} k^{*3}
    \exp \left[ 
    -\frac{1}{2}(\lambda_1 + \lambda_2 + \lambda_3)^2 
    -\frac{5}{4}\{ (\lambda_1-\lambda_2)^2+(\lambda_2-\lambda_3)^2+(\lambda_1-\lambda_3)^2 \}
    \right] \notag \\
    &\cdot\lambda_1 \lambda_2 \lambda_3 (\lambda_1-\lambda_2)(\lambda_2-\lambda_3)(\lambda_1-\lambda_3) d^3\lambda_i.
\end{align}

\section{Details of Numerical Computation \label{sec:numerics}}
\subsection{Monte Carlo Method}
In this work, we compute the nine-dimensional integral (\ref{beta_int}) by the Monte Carlo method to obtain the PBH abundance $\beta$.
The Monte Carlo method estimates an integral by randomly sampling points within the integration domain. 
Consider a one-dimensional definite integral $I = \int F(x)dx$.
If we draw $N$ random samples $x_i$ from a probability density $f(x)$ (the "sampling function", satisfying $\int f(x) dx=1$ and $f(x)> 0$), then
\begin{align}
    I = \int F(x)dx = \int \frac{F(x)}{f(x)} f(x)dx = \left\langle \frac{F(x)}{f(x)} \right\rangle \approx \frac{1}{N} \sum_i^N \frac{F(x_i)}{f(x_i)} \coloneqq I_{\rm est},
\end{align}
where the subscript $i$ labels the $i\rm th$ sampled point. 
We use $I_{\rm est}$ to estimate $I$, with relative error of order $\mathcal{O}\left(1/\sqrt{N}\right)$. 
Increasing $N$ reduces the error; furthermore, the closer the shape of $f(x)$ is to that of $F(x)$ , the smaller the error. Hence, an appropriately chosen sampling function $f(x)$ allows us to achieve a given accuracy with fewer samples, saving computational cost.

This one-dimensional Monte Carlo method could generalize straightforwardly to the nine-dimensional integral (\ref{beta_int}). 
For the convenience of sampling, we perform the following change of variables:
\begin{align}
    &\lambda_1 = u + v +w, \notag\\
    &\lambda_2 = v + w, \notag\\
    &\lambda_3 = w.
\end{align}
Under this transformation, the integration domain of (\ref{beta_int}) becomes
\begin{align}
    \nu, u, v, w &\in [0, +\infty], \notag\\
    w_1, w_2, w_3, \tilde{D}_B, \tilde{D}_c &\in [ -\infty, +\infty].
\end{align}
We denote the integral (\ref{beta_int}) as 
\begin{align}
    \beta = \int F d\nu dudvdw d^3 w_{1,2,3} d\tilde{D}_B d\tilde{D}_c,
\end{align}
and introduce a factorized sampling density 
\begin{align}
f\bigl(\nu,u,v,w,w_{1,2,3},\tilde D_B,\tilde D_C\bigr)
 = f_\nu(\nu)\,f_u(u)\,f_v(v)\,f_w(w)\,
     f_{w_{1,2,3}}^{(3)}(w_{1,2,3})\,f_{\tilde D_B}(\tilde D_B)\,f_{\tilde D_C}(\tilde D_C),
\end{align}
where each $f_*$ is in general a different function.
We then estimate
\begin{align}
    \beta_{\rm est} \coloneqq \frac{1}{N} \sum_i^N\frac{F_i}{f_i},
\end{align}
where $F_i$ and $f_i$ denote the integrand and sampling density evaluated at the $i\rm th$ sample.

\subsection{Sampling Function}
From the form of the integrand in Eq. \eqref{beta_int}, one observes:
\begin{enumerate}
    \item $F\rightarrow 0$ as $\nu, u, v, w, w_1, w_2, w_3,\tilde{D}_B, \tilde{D}_c \rightarrow +\infty$, ;
    \item $F$ contains factors $v^{1.5},u^{1.5},v^{1.5},w^{1.5}$.
\end{enumerate}

Accordingly, we adopt the following sampling distributions:
\begin{align}
&\nu \sim \Gamma(\alpha=2.5, \theta_{\nu}), \notag\\
&u \sim \Gamma(\alpha=2.5, \theta_u),
v \sim \Gamma(\alpha=2.5, \theta_v),
w \sim \Gamma(\alpha=2.5, \theta_w), \notag \\
&w_1 \sim \mathcal{N}(0, \sigma_{w_1}^2),
w_2 \sim \mathcal{N}(0, \sigma_{w_2}^2),
w_3 \sim \mathcal{N}(0, \sigma_{w_3}^2), \notag \\
&\tilde{D}_B\sim \mathcal{N}(0, \sigma_{\tilde{D}_B}^2), 
\tilde{D}_C \sim \mathcal{N}(0, \sigma_{\tilde{D}_C}^2).
\end{align}
For example, the sampling function for $\nu$ and $w_1$ read
\begin{align}
    f_\nu(\nu) &= \frac{\nu^{1.5}e^{-\nu/\theta_\nu}}{\theta_\nu^{2.5} \Gamma(2.5)}, \notag\\
    f_{w_1}(w_1) &= \frac{1}{\sqrt{2\pi}\sigma_{w_1}}\exp \left(  -\frac{w_1^2}{2\sigma_{w_1}^2}\right),
\end{align}
where the scale parameters satisfy
\begin{align}
    \theta_v &= \frac{1}{2.5} \int_0^{+\infty} \nu f(\nu)d\nu \notag \\
    \sigma_{w_1} &=  \sqrt{\frac{\pi}{2}} \int_{-\infty}^{+\infty} |w_1| f(w_1)d w_1.
\end{align}
In total there are nine tunable parameters: $\theta_\nu, \theta_u, \theta_v, \theta_w, \sigma_{w_1}, \sigma_{w_2}, \sigma_{w_3}, \sigma_{\tilde{D}_B}, \sigma_{\tilde{D}_C}$.
They should be chosen so that the sampled points concentrate in the regions that contribute most to the integral. In practice, one finds that only $\nu, u, v, w \lesssim \mathcal{O}(1)$ and $w_1, w_2, w_3, \tilde{D}_B, \tilde{D}_c \lesssim \mathcal{O}(\sigma_h^*)$ give significant contributions \footnote{One can compute the contribution of some small subregions to the integral, and it is easy to see that for subregions outside this range, the contribution is nearly zero.}.

\subsection{Optimal Parameters}
In theory, there exists an "optimal" set of parameters for which the sampling function matches the integrand shape most closely, minimizing the error. 
Taking $\theta_\nu$ as an example, suppose the shape of the sampling function is exactly the same as that of the integrand $F$, then
\begin{align}
    \theta_v = \frac{1}{2.5} \frac{\int \nu FdV}{\int FdV} 
    \approx \frac{1}{2.5} \frac{\frac{1}{N}\sum_i^N \frac{\nu_i F_i}{f_i}}{\frac{1}{N}\sum_i^N \frac{F_i}{f_i}}.
\end{align}
We estimate such optimal parameters iteratively.
In this paper, we perform several Monte Carlo trials (each with $N$ samples), and in each trial the sampling-function parameters are estimated from the previous trials.
For convenience, we define
\begin{align}
    A^{(m)} \coloneqq \frac{1}{N}\sum_i^N \frac{\nu_i F_i}{f_i},
\end{align}
where the superscript $(m)$ denotes the $m \rm th$ trial, and similarly we denote by $\beta_{\rm est}^{(m)}$ the value of $\beta_{\rm est}$ obtained in the $m \rm th$ trial.
The specific algorithm is as follows: 
\begin{enumerate}
    \item We first choose initial parameter values arbitrarily within $\theta_{\nu, u, v, w} \lesssim \mathcal{O}(1)$ and $\sigma_{w_1, w_2, w_3, D_B, D_C} \lesssim \mathcal{O}(\sigma_h^*)$, perform the first Monte Carlo trial, compute $A^{(1)},\beta_{\rm est}^{(1)}$, and set $A=A^{(1)},\beta_{\rm est}=\beta_{\rm est}^{(1)}$;
    \item In the second trial, set $\theta_\nu = \frac{1}{2.5}\frac{A} {\beta_{\rm est}}$, compute $A^{(2)},\beta_{\rm est}^{(2)}$,  and then set $A=\left(A^{(1)}+A^{(2)} \right)/2,\beta_{\rm est}=\left(\beta_{\rm est}^{(1)}+\beta_{\rm est}^{(2)} \right)/2$;
    \item In the third trial, again set $\theta_\nu = \frac{1}{2.5}\frac{A} {\beta_{\rm est}}$, compute $A^{(3)},\beta_{\rm est}^{(3)}$,  and set $A=\left(A^{(1)}+A^{(2)}+A^{(3)} \right)/3,\beta_{\rm est}=\left(\beta_{\rm est}^{(1)}+\beta_{\rm est}^{(2)}+\beta_{\rm est}^{(3)} \right)/3$.
\end{enumerate}
By repeating in this manner, as the number of trials increases, the parameter values converge toward the optimal parameters, and the computed  $\beta_{\rm est}$ approaches the true value.
In practice, the parameters for the first trial are not chosen at random: prior to the main trials, we perform preliminary trials to obtain approximate parameter values as the initial values for the main trials, thereby accelerating convergence to the optimal parameters.

Balancing computational cost and accuracy, we control the Monte Carlo integration error to about $1\%$.
For other integrals in this paper, such as Eq. \eqref{nbh_int}, we first choose appropriate sampling functions and then apply exactly the same algorithm.


%

\end{document}